\documentclass[twocolumn,showpacs,aps,pra,floatfix,nofootinbib]{revtex4-1}
\usepackage{bm}
\usepackage{amsfonts}
\usepackage{amssymb}
\usepackage{amsmath}
\usepackage{graphicx}
\usepackage{epstopdf}
\pdfoutput=1
\begin{document}

\title{Classical probability density distributions with uncertainty relations for ground states of simple non-relativistic quantum-mechanical systems}
\author{Tomasz Rado\.zycki}
\email{t.radozycki@uksw.edu.pl} \affiliation{Faculty of Mathematics and Natural Sciences, College of Sciences,
Cardinal Stefan Wyszy\'nski University, W\'oycickiego 1/3, 01-938 Warsaw, Poland}

\begin{abstract}
The probability density distributions for the ground states of certain model systems in quantum mechanics and for their classical counterparts are considered. It is shown, that classical distributions are remarkably improved by incorporating into them the Heisenberg uncertainty relation between position and momentum. Even the crude form of this incorporation makes the agreement between classical and quantum distributions unexpectedly good, except for the small area, where classical momenta are large. It is demonstrated that the slight improvement of this form, makes the classical distribution very similar to the quantum one in the whole space. The obtained results are much better than those from the WKB method. The paper is devoted to ground states, but the method applies to excited states too.
\end{abstract}
\maketitle
\section{Introduction}
\label{sintro}

It is a well known fact, that the spatial probability density distribution in quantum mechanics for low lying states entirely disagrees with the distribution found in classical mechanics. For instance the ground state wave-function of the harmonic oscillator --- and the probability density as well --- has a Gaussian character with a strong maximum around the center of the force, whereas the classical probability density acquires the largest values (or, more precisely, diverges as the inverse square root) close to the turning points. There is nothing strange in this behavior: the classical probability for finding the particle in a certain interval $[x, x+dx]$ is proportional to the time spent within it, i.e. to $dx/v(x)$. Close to the turning points the particle velocity approaches zero so the probability density infinitely increases. 

The common explanation of this discrepancy is that the quantum-mechanical ground state of certain energy $E$ (which is nonzero due to the Heisenberg uncertainty principle) does not correspond to the classical state of the same energy~\cite{burk}. The classical ground state is a state of a particle at rest (i.e. of zero energy) and, as such, is represented by the Dirac delta function $\delta(x)$, and not that describing a particle bouncing between the walls.

\begin{figure}[ht]
\centering
{\includegraphics[width=0.45\textwidth]{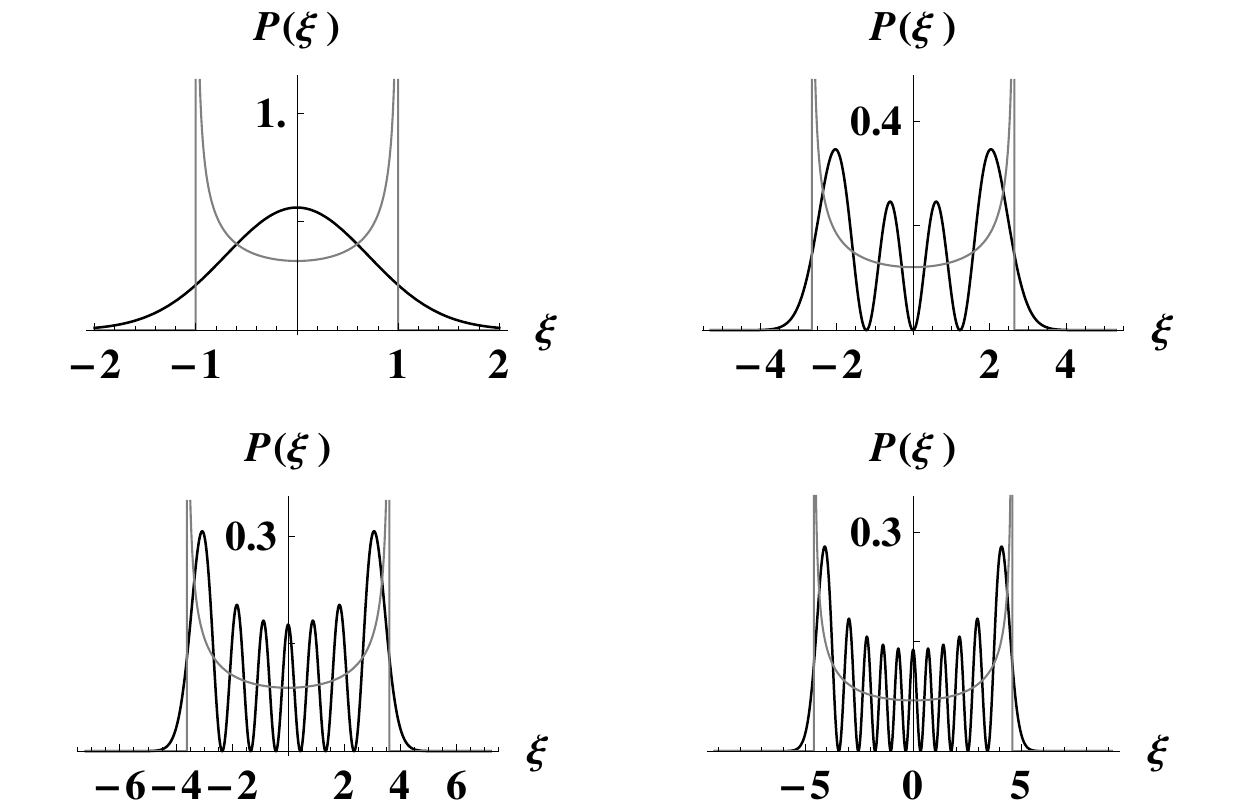}
\caption{The comparison of classical (gray lines) and quantum (black lines) probability density distributions for the harmonic oscillator. On the subsequent plots $n=0, 3, 6, 10$. $\xi$~denotes the dimensionless position defined in~(\ref{d1a}). The classical total energy on each plot is put to be equal the quantum one: $E_n=\hbar\omega(n+1/2)$.}
\label{oscold}}
\end{figure}

The disagreement between quantum and classical probability distributions is not so severe for high energy states. A well known picture of this phenomenon for the harmonic oscillator is presented in Fig. \ref{oscold}. The subsequent quantum probability density plots for increasing quantum number $n$ (drawn in black lines) more and more approach --- after the appropriate averaging --- the classical distribution (gray lines). This effect constitutes the subject of the {\em correspondence principle} \cite{bohr,hol}, which in the widely accepted although not unanimous \cite{jammer,darg} formulation states, that the predictions of quantum mechanics and classical mechanics should agree for  $n\rightarrow\infty$. Similar results are obtained for other potentials~\cite{robin}.

It was admittedly shown, in the spirit of the statistical interpretation of quantum mechanics, that the moments for canonical variables do agree in both cases~\cite{devi} , but the spatial distributions of the probability density remain dissonant. The other approach consisted in considering a localized wave-packet, representing the trapped particle, bouncing between cavity walls~~\cite{sax,brandt,bdr}. However such a wave-packet, of the size much smaller than that of the cavity, corresponds to the particle motion with a relatively high energy, certainly much larger than that of ground state. This model is, therefore, related to the situation of the last plot of Fig.~\ref{oscold}, for which the compliance between classical and quantum results does not raise any doubt, and not with the first one, we are interested in. A low-energy wave-packet should have an extent comparable to the cavity size. 

In~\cite{mil} the idea of the averaging of the classical probability density over various classical states (i.e. states with different energies) was formulated. The obtained result is in rough agreement with quantum distribution --- in particular the divergences close to the turning points are removed. However the incorporation of the uncertainty principle for the concrete classical state --- in the way suggested below --- instead of smearing the probability over different states, seems to be conceptually more justifiable and leads to much more accurate results.

The aim of this paper is then to show that the incompatibilities between probabilities in classical and quantum mechanics may be strongly reduced or even eliminated with the use of the Heisenberg uncertainty principle. It is the most essential ingredient of quantum mechanics --- some kind of its cornerstone --- and it constitutes the source of the fundamental distinctions between classical and quantum physics. Therefore it has to be inevitably included in our considerations if we want to reproduce quantum results even approximately. Below we show, that actually one can relatively easily obtain the agreement between classical and quantum probability distributions, even for the ground state, if one incorporates into the former the uncertainty relation between position and momentum in the qualitative form:
\begin{equation}
\Delta x\,\Delta p\approx \hbar.
\label{ur}
\end{equation}
No other quantum postulates are necessary. We use the classical distributions and simply smear them according to~(\ref{ur}), obtaining that way the corrected distributions.

We would like to emphasize, that throughout this paper we concentrate solely on ground states, since the agreement for highly excited states is well established and commonly accepted. In contrast, the ground state is viewed as purely quantum and it is worth specifying the conditions under which it remains in harmony with the classic one.
This relation (\ref{ur}) --- what may seem surprising while looking at the first plot of Fig.\ref{oscold} ---  turns out to be sufficient to transform the gray line into the black one with a relatively good approximation. 

The main idea of this work is formulated in the next section. In the following section we consider first the quantum and classical distributions for ground states of certain simple one-dimensional mechanical systems: the harmonic oscillator, the particle in gravitational potential (or in uniform electric field), in Morse potential and in some special potential and next try to generalize the method for the multidimensional cases: the hydrogen atom and two-dimensional asymmetric harmonic oscillator. We start with determining the classical probability distributions in these cases and then modify them by including the uncertainty principle. These modified distributions are then compared with quantum ones.

\section{The formulation of the method}
\label{formula}

In the usual approach the classical probability density $P_{Cl}(x)$ for the bound system in one dimension is defined as ratio of the time spent by a moving particle in the interval $[x, x+dx]$ to the total time $T$ needed to move between turning points (half of the period):
\begin{equation}
P_{Cl}(x)dx=\frac{dt}{T}=\frac{dx/v(x)}{T},
\label{pcl}
\end{equation}
where $v(x)$ is the velocity at a given point $x$. Close to the turning points, where the particle decelerates to zero, $P_{Cl}(x)$ diverges. On the other hand the quantum probability density $P_Q(x)$ for the ground state has a maximum in the middle of the potential cavity. These two functions substantially differ, as may be seen on the first plot of Fig.~\ref{oscold}. The main idea of the present work is to {\em slightly} modify the classical distribution in order to embody the Heisenberg uncertainty principle. 

We would like to stress that the modification we wish to introduce should be very simple. We treat it as an essential ingredient of the proposed approach. We do not want to create a sophisticated procedure to improve $P_{Cl}(x)$ in order to create the new `classical' distribution, which would be as complicated as solving Schr\"odinger equation itself,  but our purpose is to show, that when the uncertainty principle is incorporated even in the trivial and elementary manner, the discrepancy between the new probability $P(x)$ and $P_Q(x)$ becomes minor in large space area. 

The idea formulated below is based on the behavior of the quantum bouncer in the harmonic potential and in other potentials. It is known, that the high energy wave-packet is a superposition of the large spectrum of stationary oscillator states. Close to the turning points, when kinetic energy decreases, the long wavelength components are subject to the interference, which results in the reduction of tails and enhancement of the peak height~\cite{schr,brandt,lj}, similarly as the superposition of plane waves can lead to the Dirac function $\delta(x)$. For low-energy wave-packets the number of superposed states is strongly reduced and such an interference does not take place. Contrary, due to the decreasing kinetic energy de Broglie wavelength increases and the wave-packet broadens. This effect is well illustrated in~\cite{brandt}. The classical particle behaves like high-energy wave-packet. But for low lying states the uncertainty relation compensates the behavior resulting from~(\ref{pcl}).

If the classical point particle --- according to Ehrenfest idea~\cite{eh} --- is to be promoted to the bouncing wave packet, then its position should become unknown up to a couple of the de Broglie wavelengths:
\begin{equation}
\Delta x\sim n \lambda_{dB}= n\,\frac{h}{p},
\label{db}
\end{equation}
where $n$ is a certain positive number (not necessarily an integer). 
To facilitate our further analysis and to allow for comparison of the results in different models, in each of the cases considered in the following section we define the dimensionless position $\xi$ and momentum $\eta$ simultaneously eliminating the dimensional constants present in every model. For these variables the uncertainty principle~(\ref{ur}) reads
\begin{equation}
\Delta\xi\,\Delta\eta\approx 1.
\label{urm}
\end{equation}

Let us now assume that $\Delta \eta$ is maximally of order of $\eta$ itself, by putting $\Delta\eta=\kappa|\eta|$, where $\kappa\leq 2$ is a certain parameter to be fixed later. The `$2$' comes from the utmost classical values of momentum at each point: from $-|\eta|$ to $+|\eta|$. It will turn out, that the optimal value of this parameter and, what should especially be pointed out, {\em common} for all considered cases (except that in two dimensions) is $\kappa=1.7$. Now 
upon~(\ref{db}) we can put 
\begin{equation}
\Delta\xi= \frac{1}{\kappa|\eta|}.
\label{uncpos}
\end{equation}
Since inside a cavity the momentum $\eta$  is in general $\xi$-dependent, the same refers to the indefiniteness of $\xi$, which will change from point to point.

How to incorporate~(\ref{urm}) and~(\ref{uncpos}) into classical probability distribution? We achieve this by assuming that the value $P_{Cl}(\xi)$ is not attributed to the single point $\xi$ but to the whole `cell' of the spread $\Delta\xi$ around it. Therefore, to the corrected probability density at a certain point do contribute all cells (potentially with different weights), that contain it.

We introduce below a certain function $\Phi(\xi,\xi')$, which specifies in what way the value $P_{Cl}(\xi')$ is distributed within such an elementary cell. The most natural will be to choose it in the Gaussian form, but we start with the trivial case of a properly normalized constant in the whole interval (throughout the paper this interval will be called `the uncertainty cell'):
\begin{equation}
D_{\xi'}=[\xi_-, \xi_+],
\label{interval}
\end{equation}
where
\begin{equation}
\xi_\pm=\xi'\pm\frac{1}{2\kappa|\eta(\xi')|}=\xi'\pm \frac{\Delta\xi}{2}.
\label{xipm}
\end{equation}
The size of this interval changes together with $\xi'$ and becomes larger close to the turning points. 
We then first define the function $\phi(\xi,\xi')$
\begin{equation}
\phi(\xi,\xi')=(\Delta\xi)^{-1}\chi_{\xi'}(\xi)=\kappa|\eta(\xi')|\chi_{\xi'}(\xi),
\label{phi}
\end{equation}
where $\chi_{\xi'}$ is the characteristic function of the set $D_{\xi'}$, and next
\begin{equation}
\Phi(\xi,\xi')=\phi(\xi,\xi') P_{Cl}(\xi').
\label{pphi}
\end{equation}
 It obviously satisfies
\begin{equation}
\int\limits_{\xi_-}^{\xi_+} \Phi(\xi,\xi')d\xi=P_{Cl}(\xi').
\label{normphi}
\end{equation}

The new probability distribution is now created as the sum (or rather integral) over contributions coming from each cell (the integral over $\xi'$ is in fact the integral over different cells)
\begin{equation}
P(\xi)=\int\limits_{\xi_{min}}^{\xi_{max}}\Phi(\xi,\xi')d\xi',
\label{prob}
\end{equation}
where $\xi_{min}$ and $\xi_{max}$ denote the turning points of the bouncing particle. This formula constitutes our main instrument in the following section. The function $\Phi(\xi,\xi')$ has an interesting property: integrated over first argument gives  the old probability distribution $P_{Cl}$, while integration over the second one leads to the improved distribution $P$.

The choice of the function $\phi$ as given by~(\ref{phi}) is the simplest one, but as we shall see below, already in this form it is sufficient to reproduce relatively well the quantum probability distribution of the ground states, except for certain small regions. Our goal is to use, however, the Gaussian form of $\phi(\xi,\xi')$, for which the results are still better. 

\section{Numerical results}
\label{num}

\subsection{One-dimensional systems}
\label{oned}

In this section we apply our formula~(\ref{prob}) to a couple of simple one-dimensional quantum-mechanical systems. We begin with the harmonic oscillator and continue with other examples for which the quantum solutions are well known.

\subsubsection{Harmonic oscillator}
\label{osci}

Let us consider classical one-dimensional system described by the Hamiltonian
\begin{equation}
{\cal H}=\frac{p^2}{2m}+\frac{m\omega^2x^2}{2}.
\label{hosc}
\end{equation}
The ground state energy of the quantum version of the system is $E=\hbar\omega/2$. We will consider the classical motion of the same energy (and not of the zero energy!) and compare $P_Q(x)$ and $P(x)$.

Let us introduce dimensionless variables
\begin{subequations}\label{d1}
\begin{align}
\xi&=\sqrt{\frac{m\omega}{\hbar}}\,x,\;\;\;\;\; \eta=\frac{p}{\sqrt{m\hbar\omega}}, \label{d1a}\\
\tau&=\omega t,\;\;\;\;\; {\cal E}=\frac{E}{\hbar\omega}.\label{d1b}
\end{align}
\end{subequations}

The oscillator equation for the movement of energy $E$ in the new variables reads
\begin{equation}
\frac{\xi^2}{2}+\frac{\eta^2}{2}={\cal E},
\label{eosc}
\end{equation}
with
\begin{equation}
\frac{d\xi}{d\tau}=\eta,
\label{dke}
\end{equation}
and turning points are
\begin{equation}
\xi_{min}=-\sqrt{2{\cal E}},\;\;\;\;\; \xi_{max}=\sqrt{2{\cal E}}.
\label{turnosc}
\end{equation}
At a certain point $\xi$, the corresponding momentum is obviously equal to
\begin{equation}
\eta(\xi)=\pm\sqrt{2{\cal E}-\xi^2}.
\label{momosc}
\end{equation}
Since the dimensionless period is equal to $2\pi$, according to~(\ref{pcl}) we have
\begin{equation}
P_{Cl}(\xi')=\frac{1}{\pi|\eta(\xi')|},
\label{plosc}
\end{equation}
and we are in the position to write down the function $\Phi(\xi,\xi')$:
\begin{equation}
\Phi(\xi,\xi')=\frac{1}{\pi|\eta(\xi')|}\, \kappa|\eta(\xi')|\chi_{\xi'}(\xi)=\frac{\kappa}{\pi}\, \chi_{\xi'}(\xi).
\label{phiosc}
\end{equation}

For the state in question (both classical and quantum) we put ${\cal E}=1/2$ and then the modified probability distribution will be given by
\begin{equation}
P(\xi)=\frac{\kappa}{\pi}\int\limits_{-1}^{1}\chi_{\xi'}(\xi)d\xi',
\label{Posc}
\end{equation}
with $D_{\xi'}$ defined by~(\ref{interval}) and $|\eta(\xi')|=\sqrt{1-\xi'^2}$.
This result should be compared with the quantum probability distribution for ground state, which is well known~\cite{schiff}, and has the form:
\begin{equation}
P_Q(\xi)=\frac{1}{\sqrt{\pi}}\, e^{-\xi^2}.
\label{pqosc}
\end{equation}

\begin{figure}[ht]
\centering
{\includegraphics[width=0.45\textwidth]{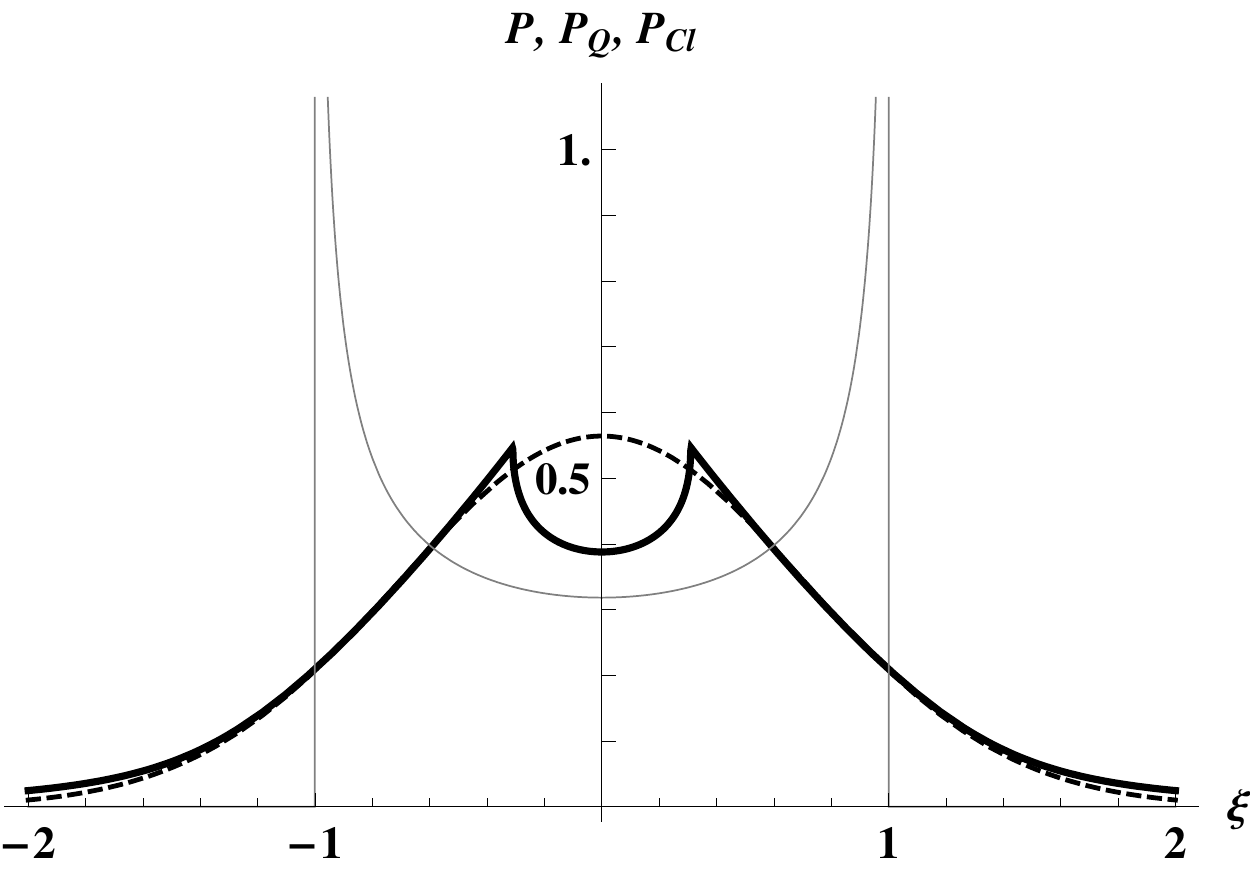}
\caption{Plot of the modified `classical' ground-state distribution $P(\xi)$ defined by~(\ref{plosc}) drawn with solid line. The parameter $\kappa$ is chosen to be $1.7$. For comparison $P_Q(\xi)$ -- the quantum distribution (dashed line) and $P_{Cl}(\xi)$ -- the classical distribution (gray line) are shown.}
\label{osc1}}
\end{figure}

In Fig.~\ref{osc1} the modified probability distribution $P(x)$ is plotted with solid line. Except for the middle of the cavity it perfectly agrees with the quantum distribution $P_Q(\xi)$ plotted with dashed line. Not only the singularities of the classical distributions have been cured, but also the penetration of the classically forbidden region has been reconstructed. As it has already been advertised it is noticeable that the best fit here and for {\em all} following curves corresponds to the value $\kappa=1.7$. The unmodified classical distribution $P_{Cl}(\xi)$ for the same energy is shown in gray color.

\begin{figure}[ht]
\centering
{\includegraphics[width=0.45\textwidth]{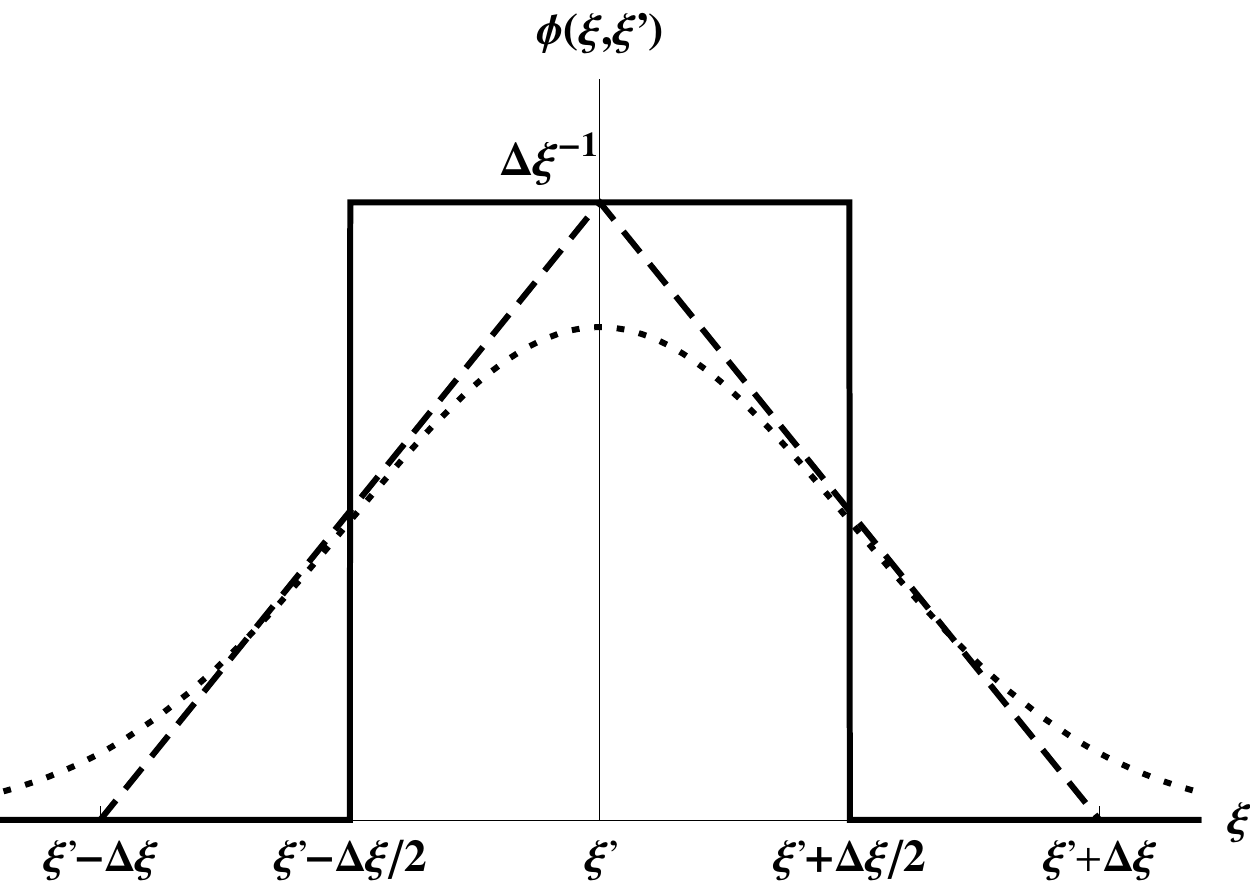}
\caption{The various possible forms of the function $\phi(\xi,\xi')$: the step behavior (solid line), the triangle behavior (dashed line) and the Gaussian behavior (dotted line). The area under all curves is normalized to unity.}
\label{heisen}}
\end{figure}

This result demonstrates the fundamental role played by uncertainty principle. The inclusion of it even in the rough form is enough to obtain nearly perfect agreement in large areas. The visible deviation from the quantum result in the center may be explained by the coarse form of~(\ref{phi}), which is a constant within the interval $D_{\xi'}$. For small values of $\xi$, where the momentum $\eta(\xi)$ becomes relatively large, its indefiniteness is overestimated. This, via~(\ref{uncpos}), results in a relatively narrow interval $D_{\xi'}$, thus conformed to the classical case.

To verify this observation, we modify below the behavior of $\phi(\xi,\xi')$ to improve its sharp shape~(\ref{phi}) and distribute $P_{Cl}(\xi')$ more smoothly inside the uncertainty cell. In Fig.~\ref{heisen} we show simple alternatives: the triangle and the Gaussian behavior. We choose two different forms to verify, how sensitive to them the probability distribution $P(\xi)$ is.

In the first case instead of~(\ref{phi}), we have
\begin{align}
\phi(\xi,\xi')&=\bar{\chi}_{\xi'}(\xi)(\Delta\xi)^{-2}(\Delta\xi-|\xi-\xi'|)\nonumber\\
&=\bar{\chi}_{\xi'}(\xi)(\kappa|\eta(\xi')|)^2\left(\frac{1}{\kappa|\eta(\xi')|}-|\xi-\xi'|\right),
\label{phi2}
\end{align}
where $\bar{\chi}_{\xi'}(\xi)$ is the characteristic function of the interval $[\xi_-,\xi_+]$ but now with modified $\xi_\pm$:
\begin{equation}
\xi_\pm=\xi'\pm\frac{1}{\kappa|\eta(\xi')|}=\xi'\pm \Delta\xi.
\label{xipm2}
\end{equation}

\begin{figure}[ht]
\centering
{\includegraphics[width=0.45\textwidth]{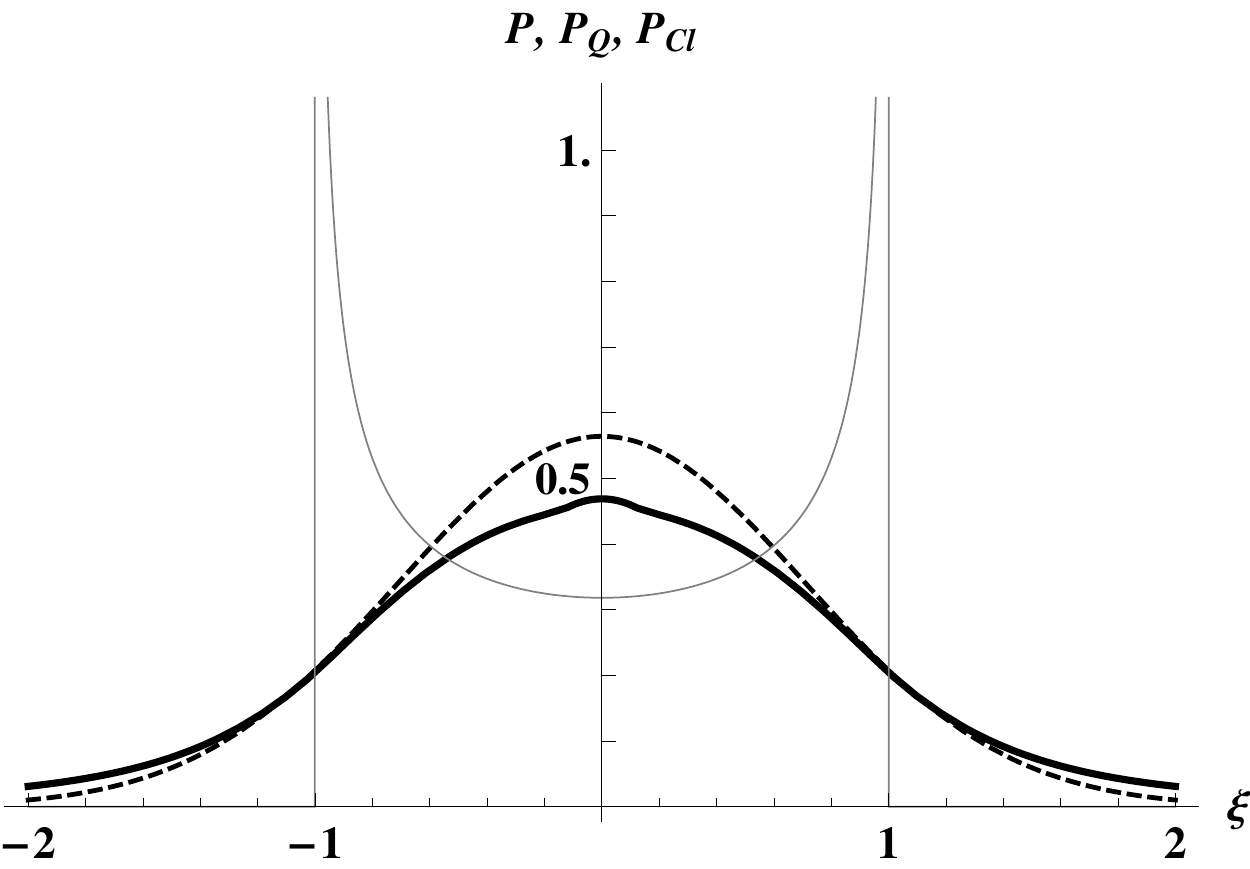}
\caption{Same as Fig.~\ref{osc1}, but with $P(x)$ found with the use of~(\ref{phi2}).}
\label{osc2}}
\end{figure}

For $\Phi(\xi,\xi')$ we obtain
\begin{equation}
\Phi(\xi,\xi')=\frac{\kappa^2|\eta(\xi')|}{\pi}\,\chi_{\xi'}(\xi)\left(\frac{1}{\kappa|\eta(\xi')|}-|\xi-\xi'|\right),
\label{Posc2}
\end{equation}
and again $P(x)$ will be found using formula~(\ref{prob}). The result is plotted in Fig.~\ref{osc2}. As one can see the agreement between quantum and modified classical case is strongly improved and may be considered satisfactory.

\begin{figure}[ht]
\centering
{\includegraphics[width=0.45\textwidth]{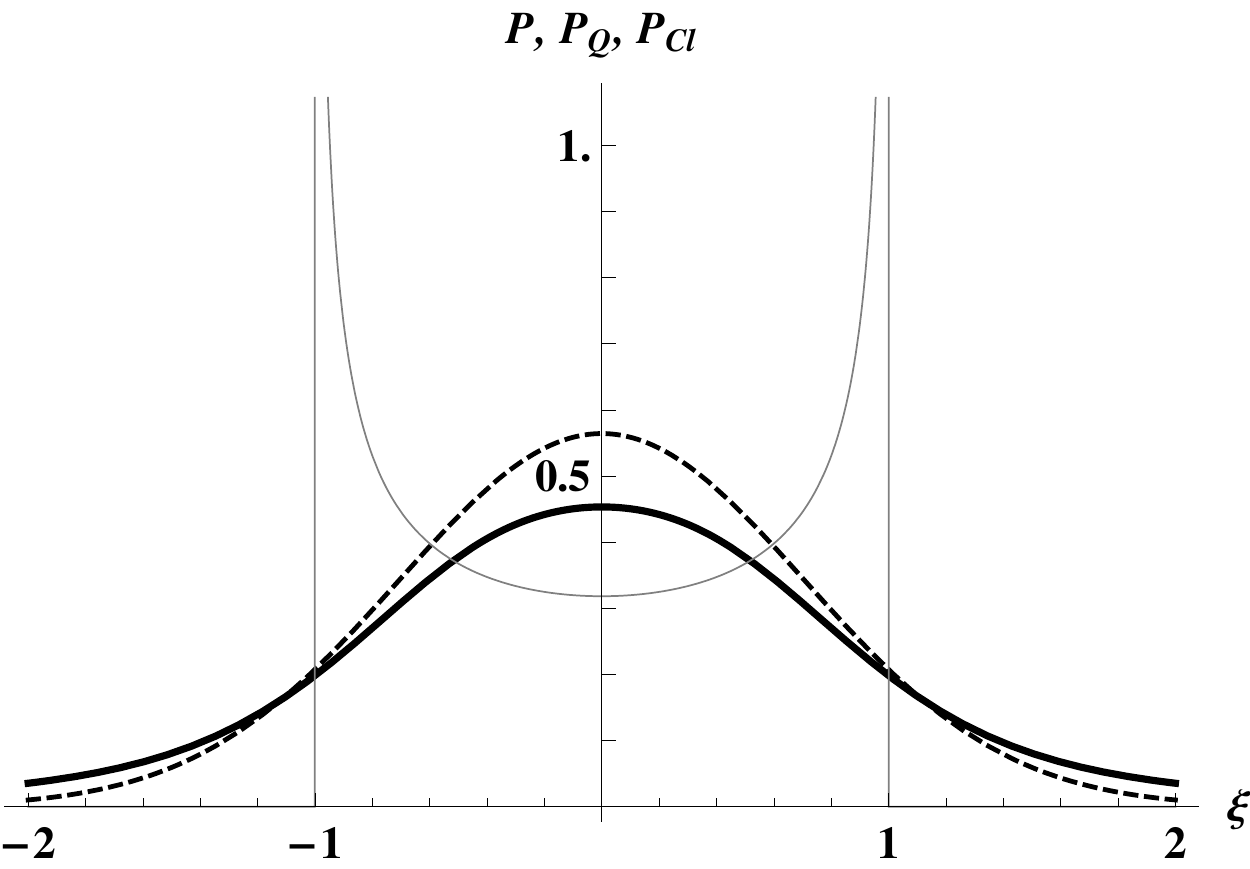}
\caption{Same as Fig.~\ref{osc2}, but with $P(x)$ found with the use of~(\ref{phi3}).}
\label{osc3}}
\end{figure}

As a Gaussian shape of $\phi(\xi,\xi')$ we in turn take
\begin{align}
\phi(\xi,\xi')=\sqrt{\frac{2}{\pi}}(\Delta\xi)^{-1}e^{-2(\xi-\xi')^2/(\Delta\xi)^2}\nonumber\\
\sqrt{\frac{2}{\pi}}\,\kappa|\eta(\xi')| e^{-2\kappa^2\eta(\xi')^2(\xi-\xi')^2},
\label{phi3}
\end{align}
which gives
\begin{equation}
\Phi(\xi,\xi')=\sqrt{\frac{2}{\pi^3}}\,\kappa e^{-2\kappa^2\eta(\xi')^2(\xi-\xi')^2}.
\label{Posc3}
\end{equation}

It may be easily verified, that the condition~(\ref{normphi}) is satisfied, with the appropriate change of the integration limits:
\begin{equation}
\int\limits_{-\infty}^{\infty} \Phi(\xi,\xi')d\xi=P_{Cl}(\xi').
\label{normphi3}
\end{equation}

The probability distribution obtained according to~(\ref{prob}) is plotted in Fig.~\ref{osc3}. The agreement of this curve with the quantum result is again very good. It is also noteworthy that the difference between plots in Figs~\ref{osc2} and~\ref{osc3} is hardly visible. This suggests that the details of the form of the function $\phi(\xi,\xi')$ are inessential. The truly important is the inclusion of the uncertainty principle itself and the size of the uncertainty cell.

\begin{figure}[ht]
\centering
{\includegraphics[width=0.45\textwidth]{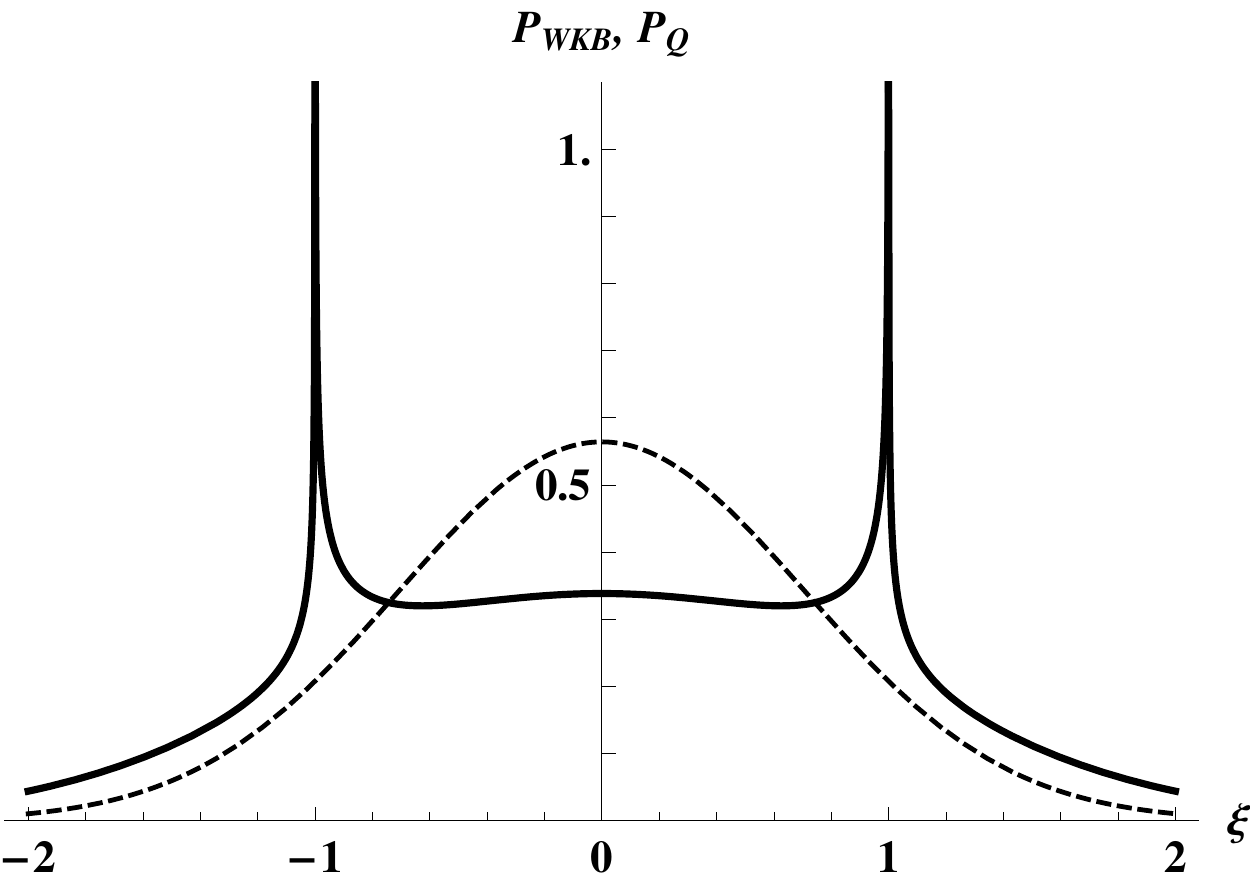}
\caption{Comparison of the WKB probability density (solid line) and the quantum one (dashed line) for the ground state harmonic oscillator.}
\label{oscWKB}}
\end{figure}

Our results even in the simplest case presented in Fig.~\ref{osc1} are much more accurate than those obtained from the WKB method~\cite{wen, kra, bri}. The comparison of the quasi-classical probability density and the quantum one is done in Fig.~\ref{oscWKB}. It is known, however, that the WKB method fails close to the turning points, as well as for low lying states~\cite{schiff}. Moreover this method transforms the quantum distribution into the (quasi) classical one rather than conversely.

\begin{figure}[ht]
\centering
{\includegraphics[width=0.5\textwidth]{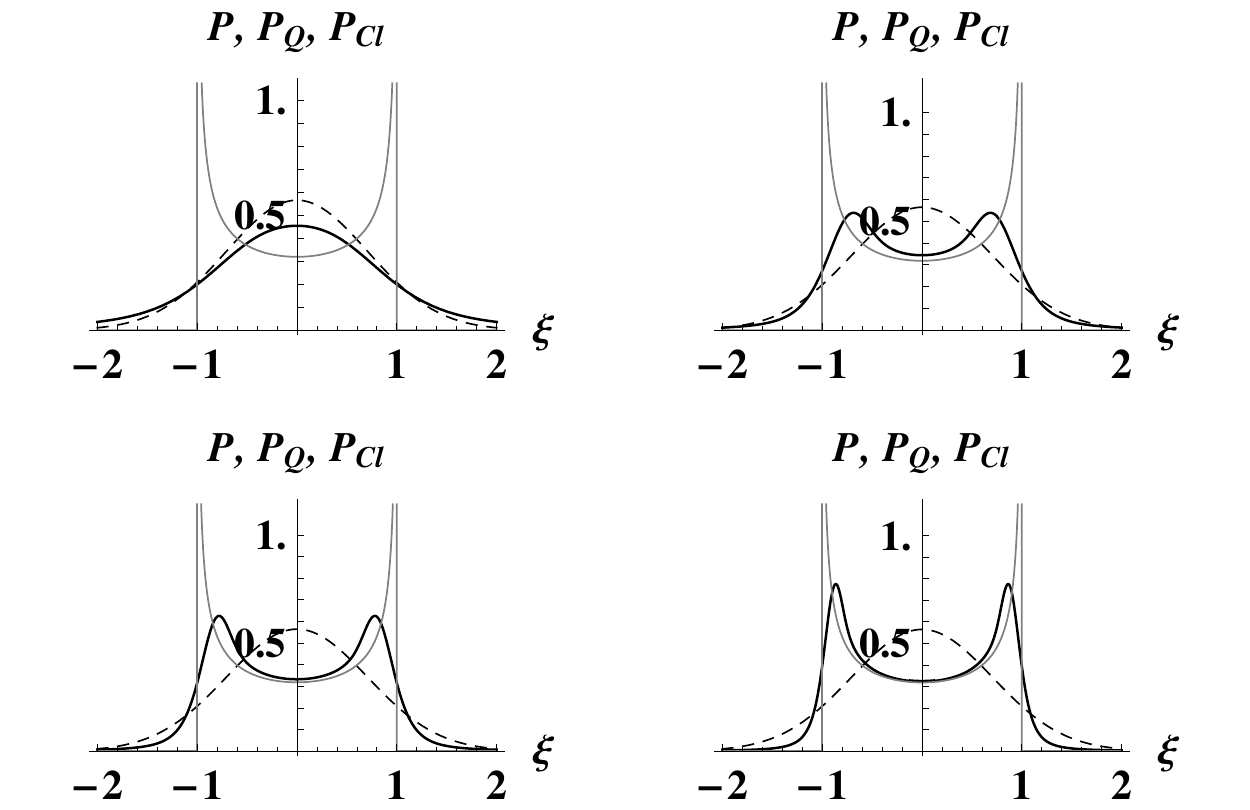}
\caption{The dependence of the probability density distribution $P(x)$ of the value of parameter $\kappa$. The successive plots are performed for $\kappa=1.7, 6, 10, 20$.}
\label{osc3r}}
\end{figure}

The parameter $\kappa$ in our approach may serve to mediate between the classical and the quantum case. Smaller values correspond to large uncertainty in position, whereas for bigger ones the uncertainty decreases, which should lead to reproducing the classical result. From Fig.~\ref{osc3r} one sees, that it is really the case. The subsequent plots are drawn for increasing values of $\kappa$ (from $1.7$ to $20$), actually approaching more and more the classical distribution. For these plots the function $\Phi(\xi,\xi')$ was taken in the Gaussian form~(\ref{Posc3}), but similar results may be obtained for~(\ref{phiosc}) and~(\ref{Posc2}). It was mentioned that the optimal value is $\kappa=1.7$.

\begin{figure}[ht]
\centering
{\includegraphics[width=0.45\textwidth]{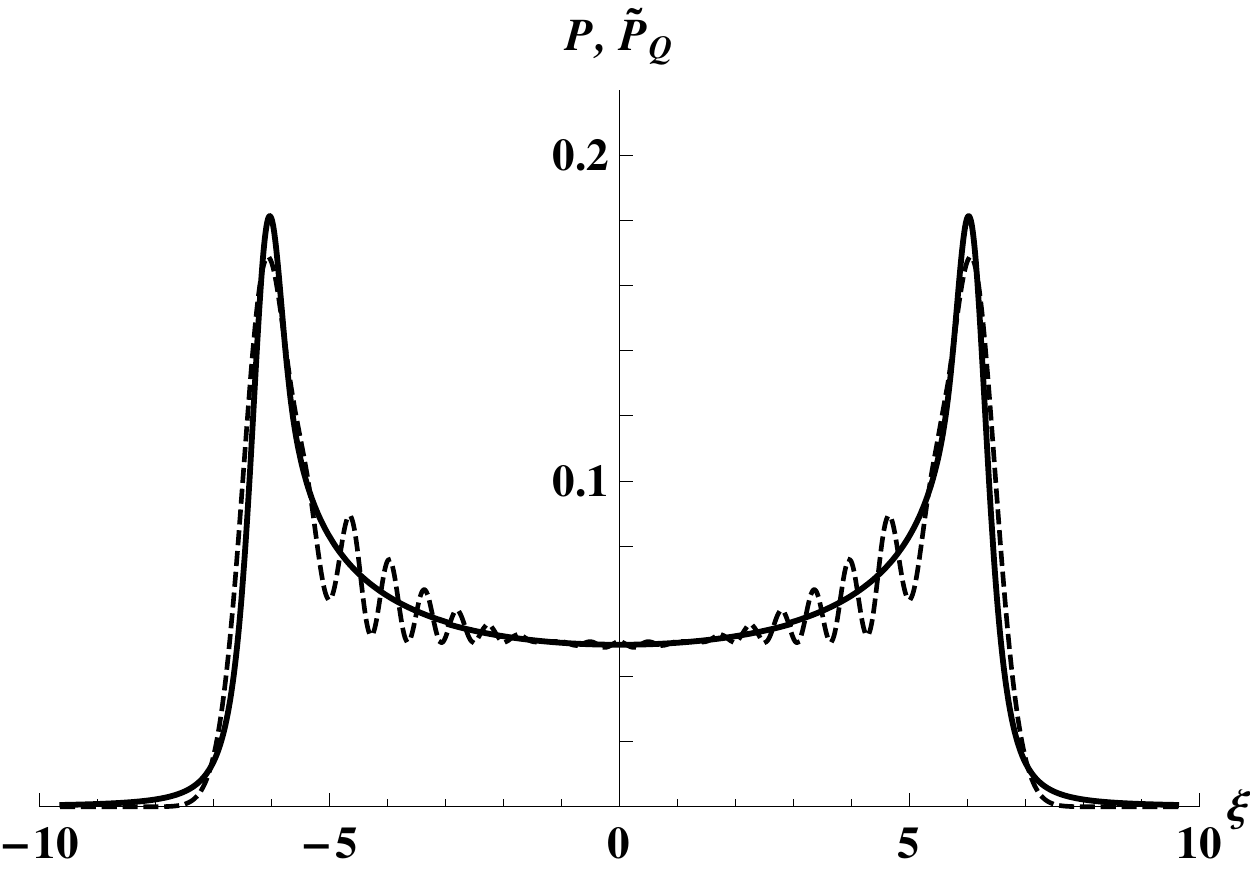}
\caption{The comparison of $P(\xi)$ (solid line) and $\tilde{P}_Q(\xi)$ (dashed line) for excited oscillator state of $n=20$. The averaging length $d$ was arbitrarily chosen as equal to $1$. The plot for $P(\xi)$ was performed with the use of~(\ref{phi3}).}
\label{oschigh3}}
\end{figure}
 
As it was told in the introduction, this paper is devoted to the ground states. However we would like to emphasize, that our formula~(\ref{prob}) exhibits the correct behavior for highly excited states too. We will not analyze this point here but merely compare in Fig.~\ref{oschigh3} the plot of $P(\xi)$ for large $n$ and the quantum distribution $\tilde{P}_Q(\xi)$, obtained by the appropriate averaging:
\begin{equation}
\tilde{P}_Q(\xi)=\frac{1}{d}\,\int\limits_{\xi-d/2}^{\xi+d/2} P_Q(\zeta) d\zeta,
\label{ave}
\end{equation} 
where $d$ is a certain averaging length. The agreement is almost perfect.
 
In the following subsections we will try to confirm these results for other quantum-mechanical systems. 

\subsubsection{Particle in gravitational field}
\label{gravi}

Let us now consider the second model, which will constitute the particle in the potential
\begin{equation}
V(x)=\left\{\begin{array}{lcc} F x, &\mathrm{for}& x\geq 0,\\ \infty, &\mathrm{for}& x<0,\end{array}\right.
\label{potgra}
\end{equation}
with $F>0$, which may be called `the quantum bouncer'~\cite{gibbs}. 
It can describe the freely falling particle in gravitational field in which case $F=mg$, with the elastic barrier for $x=0$, or a charged particle in uniform electric field. This problem is not purely academic, since such kind of bound states for neutrons in the gravitational field of the Earth has been observed~\cite{nes,liu,pedram}.

The Hamiltonian of the system is given by
\begin{equation}
{\cal H}=\frac{p^2}{2m} + mgx,
\label{hgrav}
\end{equation}
and the classical motion is limited between the initial height $H$ and $x=0$.

Introducing the dimensionless parameters in the following way
\begin{subequations}\label{d2}
\begin{align}
\xi&=\left(\frac{m^2g}{\hbar^2}\right)^{1/3}\!\!\!x,\;\;\;\;\; \eta=(m^2g\hbar)^{-1/3}p, \label{d2a}\\
\tau&=\left(\frac{
mg^2}{\hbar}\right)^{1/3}\!\!\! t,\;\;\;\;\; {\cal E}=(mg^2\hbar^2)^{-1/3}E.\label{d2b}
\end{align}
\end{subequations}
we can rewrite the energy conservation equation in the form
\begin{equation}
\frac{\eta^2}{2}+\xi={\cal E},
\label{paragra}
\end{equation}
where energy ${\cal E}$ is nothing else but the dimensionless initial height
\begin{equation}
{\cal E}=\left(\frac{m^2g}{\hbar^2}\right)^{1/3}\!\!\!H.
\label{cale}
\end{equation}
In these new variables the relations~(\ref{urm}) as well as~(\ref{dke}) hold. The turning points naturally are
\begin{equation}
\xi_{min}=0,\;\;\;\; \xi_{max}={\cal E}
\label{}
\end{equation}

To write down the formula for $P_{Cl}(\xi)$ we need $T$ i.e. half of the period of the oscillatory motion. This value is well known from school physics to be $T=\sqrt{2H/g}$, which in the dimensionless variables simply is $\sqrt{2\xi_{max}}$. We obtain then
\begin{equation}
P_{Cl}(\xi)=\frac{1}{\sqrt{2\xi_{max}}\,|\eta(\xi)|},
\label{pclgra}
\end{equation}
with $|\eta(\xi)|=\sqrt{2({\cal E}-\xi)}$.

Let us now construct $P(\xi)$ for the uniform distribution of $P_{Cl}(\xi')$ within an uncertainty cell. We cannot unconsciously take for the function $\phi(\xi,\xi')$ the form~(\ref{phi}) chosen in the case of the harmonic oscillator, since the region $\xi<0$ is now excluded from the penetration by the particle in both classical and quantum cases. Therefore, if for certain value $\xi'$ it turns out that $\xi_-<0$, then in the definition~(\ref{interval}) of $D_{\xi'}$, we abruptly put $\xi_-=0$. One can say, that we cut off the protruding portion of the interval. This of course must be accompanied by  the adjustment of the normalization constant, so as to maintain the condition~(\ref{normphi}). Formally we define the quantity $L(\xi')$ (i.e. length of the interval $D_{\xi'}$) as
\begin{equation}
L(\xi')=\left\{\begin{array}{lcl} \Delta\xi, &\mathrm{if} & \xi'-\Delta\xi/2>0,\\
\xi'+\Delta\xi/2, &\mathrm{if} & \xi'-\Delta\xi/2\leq 0.\end{array}\right.
\label{Lgra}
\end{equation}
This allows us to write the function $\phi(\xi,\xi')$ in the form
\begin{equation}
\phi(\xi,\xi')=L(\xi')^{-1}\, \chi_{\xi'}(\xi)\Theta(\xi),
\label{phigrav}
\end{equation}
where $\Theta(\xi)$ is the Heaviside step function. Now $P(\xi)$ may be found in the standard way, with the use of formulas~(\ref{pphi}), (\ref{prob}) and~(\ref{phigrav}).

In order to guarantee that the classical motion correspond to the quantum-mechanical ground state, the value of ${\cal E}$ (or equivalently of height $H$) will be chosen to be minus the first zero of the Airy $\mathrm{Ai}$ function~\cite{abra}, which simultaneously is the ground state energy got from the appropriate Schr\"odinger equation~\cite{flugge,berb,yoder}:
\begin{equation}
{\cal E}=\xi_{max}\approx 2.33811.
\label{first0}
\end{equation}
The quantum probability distribution is given by
\begin{equation}
P_Q(\xi)=N_q[\mathrm{Ai}(2^{1/3}(\xi -\xi_{max}))]^2.
\label{pqgra}
\end{equation}
where $N_q$ is the normalization constant and can be found numerically as $N_q\approx 2.5624$.

\begin{figure}[ht]
\centering
{\includegraphics[width=0.45\textwidth]{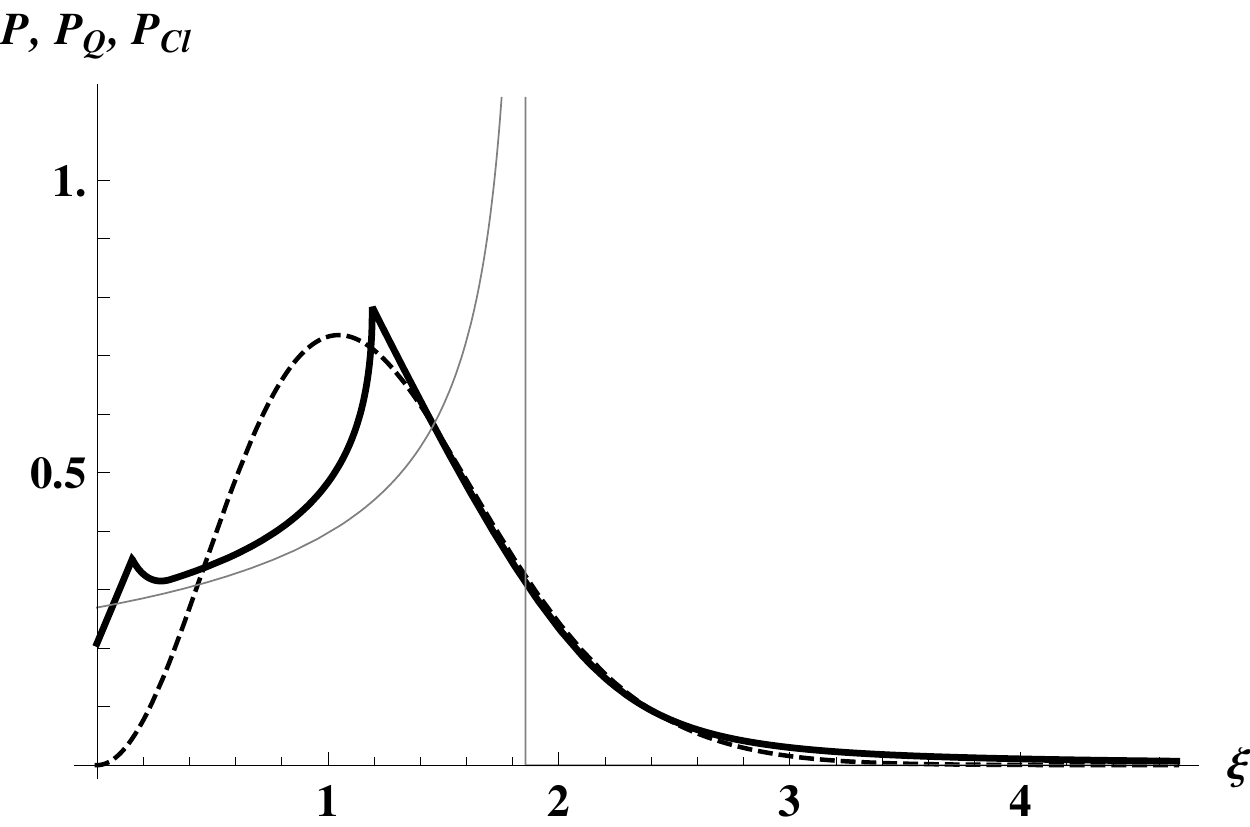}
\caption{Same as Fig.~\ref{osc1}, but for the system defined by~(\ref{hgrav}). The function $\phi(\xi,\xi')$ is given by~(\ref{phigrav}). The parameter $\kappa$ is equal to $1.7$.}
\label{grav1}}
\end{figure}

The results are plotted in Fig.~\ref{grav1}. As before $P_{Cl}(\xi)$ is drawn in  gray line. We see again the significant improvement of the classical distribution after having applied the proposed procedure. For larger values of $\xi$ the agreement between $P(\xi)$ and $P_Q(\xi)$ is perfect. The deviations for smaller values of $\xi$ are of the similar origin as those in Fig.~\ref{osc1}.

If we wish to ameliorate the behavior of $P(\xi)$ for large values of $\eta$, i.e. close to the earth surface, we can modify the function $\phi(\xi,\xi')$ in the way analogous to that of the previous subsection and shown in Fig.~\ref{heisen}. Because of the `rigid wall' at $\xi=0$, we have to, however, make some adjustments similar to that, which led to~(\ref{phigrav}) and which restrict the accessible space to $\xi>0$. The normalization has to be properly corrected as well in order to ensure~(\ref{normphi}). Let us first define
\begin{equation}
L(\xi')=\left\{\begin{array}{lcl} \Delta\xi^2, &\mathrm{if} & \xi'-\Delta\xi/2>0,\\
\Delta\xi^2-(\Delta\xi/2-\xi')^2, &\mathrm{if} & \xi'-\Delta\xi/2\leq 0.\end{array}\right.
\label{Lgra2}
\end{equation}

\begin{figure}[ht]
\centering
{\includegraphics[width=0.45\textwidth]{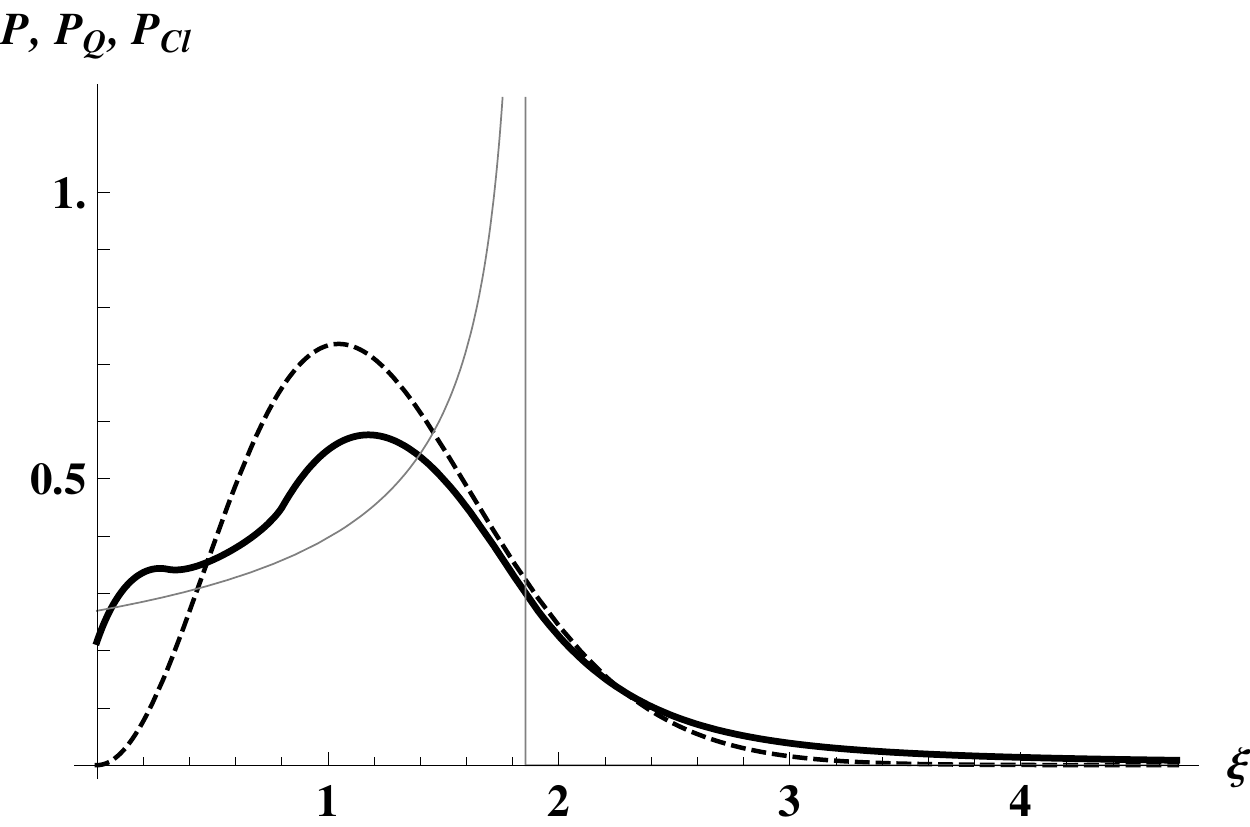}
\caption{Same as Fig.~\ref{grav1} but obtained with the use of~(\ref{phigrav2}).}
\label{grav2}}
\end{figure}

\begin{figure}[ht]
\centering
{\includegraphics[width=0.45\textwidth]{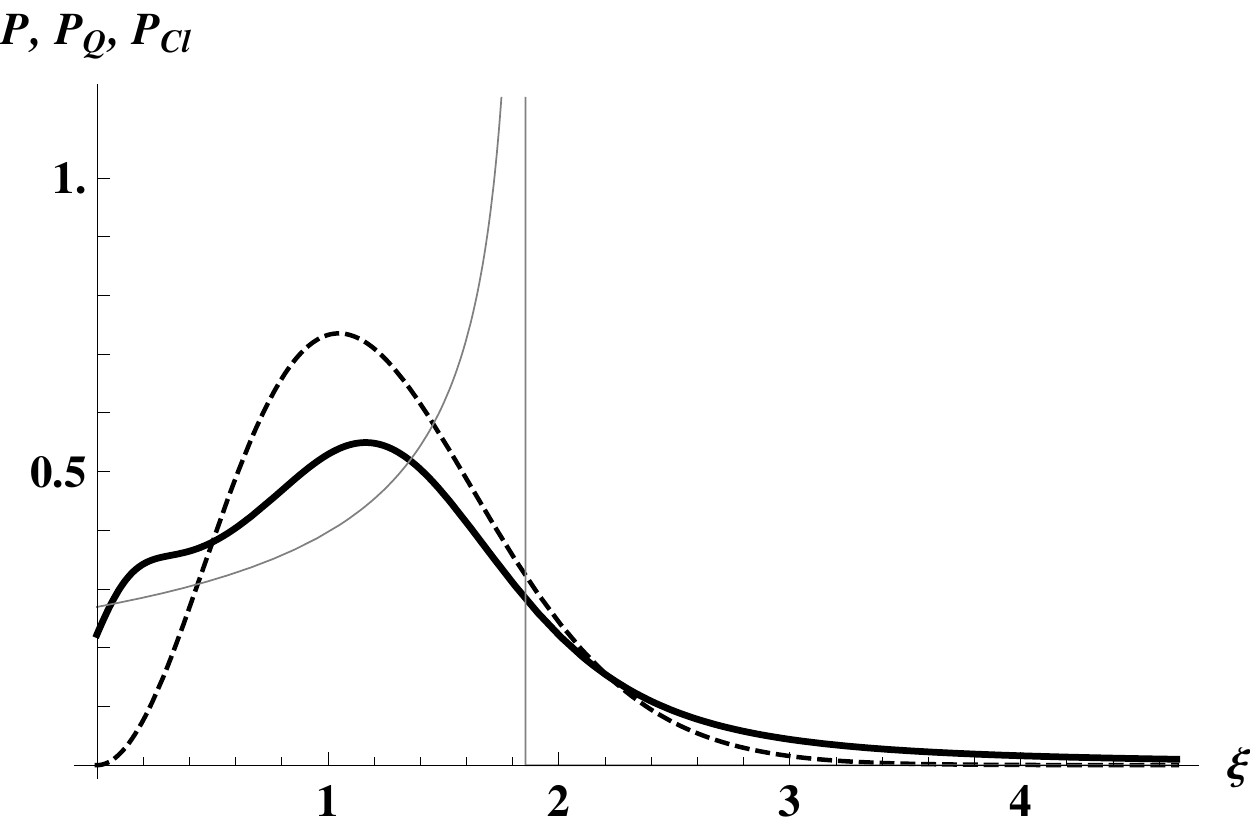}
\caption{Same as Fig.~\ref{grav1} but obtained with the use of~(\ref{phigrav3}).}
\label{grav3}}
\end{figure}

In place of~(\ref{phigrav}) we will now have
\begin{equation}
\phi(\xi,\xi')=L(\xi')^{-1}\Theta(\xi)\chi_{\xi'}(\xi)(\Delta\xi-|\xi-\xi'|),
\label{phigrav2}
\end{equation}
where $\Theta(\xi)$ is the Heaviside step function. This leads to $P(\xi)$ shown in Fig.~\ref{grav2}. Almost identical results are obtained with the Gaussian form of $\phi(\xi,\xi')$ cut from the left at $\xi=0$:
\begin{equation}
\phi(\xi,\xi')=C(\xi') \Theta(\xi)e^{-2(\xi-\xi')^2/(\Delta\xi)^2},
\label{phigrav3}
\end{equation}
where
\begin{equation}
C(\xi')=\frac{2\sqrt{2}}{\sqrt{\pi}\,\Delta\xi(1-\mathrm{erf}(-\sqrt{2}\,\xi'/\Delta\xi))},
\label{nc}
\end{equation}
and $\mathrm{erf}(x)$ is the error function. The obtained plot of $P(\xi)$ is presented in Fig.~\ref{grav3}. It is visible that the distributions are not sensitive to the details of $\phi$. The same observation was made for the harmonic oscillator

The reader may point out to the pronounced discrepancy between $P(\xi)$ and $P_Q(\xi)$ for small $\xi$. However it is connected with the unphysical notion of the ideal rigid wall. None wall is perfectly rigid. The improvement of the behavior close to such a wall can be achieved if one assumes the more smooth form of $\phi(\xi,\xi')$ on this border as for instance asymmetrically squeezed Gaussian function. This would be, however, at odds with our main idea of the only {\em slight} adjustment of the classical distribution. Such discrepancies do not arise for more physical potentials than the rigid wall in~(\ref{potgra}).

\subsubsection{Morse potential}
\label{morse}

The next example we are going to consider is the Morse potential, which constitutes a relatively good approximation for describing the vibrational degrees of freedom of a diatomic molecule~\cite{morse,atkins}. It is then a really important case of some practical significance. The corresponding Hamiltonian may be given the form
\begin{equation}
{\cal H}= \frac{p^2}{2m}+V_0\left(e^{-2x/d}-2e^{-x/d}\right),
\label{morsepot}
\end{equation}
where $V_0,d>0$. The dimensionless variables in this case are
\begin{subequations}\label{mordim}
\begin{align}
\xi&=\frac{x}{d}\, ,\;\;\;\;\; \eta=\frac{d}{\hbar}\, p\label{mordim1}\\
\tau&=\frac{\hbar}{md^2}\, t,\;\;\;\;\; {\cal E}=\frac{d^2m}{\hbar^2}\,E,\label{mordim2}
\end{align}
\end{subequations}
and in consequence the relations~(\ref{urm}) and~(\ref{dke}) are satisfied. In these variables the equation~(\ref{morsepot}) becomes simply
\begin{equation}
\frac{\eta^2}{2}+\beta(e^{-2\xi}-2e^{-\xi})={\cal E},
\label{m2}
\end{equation}
where we introduced the additional dimensionless constant 
\begin{equation}
\beta=\frac{d^2m V_0}{\hbar^2}\,.
\label{beta}
\end{equation}
For the momentum we get
\begin{equation}
|\eta(\xi)|=\sqrt{2({\cal E}-\beta(e^{-2\xi}-2e^{-\xi}))}
\label{morseeta}
\end{equation}

Solving~(\ref{m2}) for $\eta=0$, we easily find the turning points (${\cal E}<0$ for bound states):
\begin{subequations}\label{xmm}
\begin{align}
\xi_{min}&=-\ln(1+\sqrt{1+{\cal E}/\beta}),\label{xmm1}\\
\xi_{max}&=-\ln(1-\sqrt{1+{\cal E}/\beta}).\label{xmm2}
\end{align}
\end{subequations}

To write down the formula~(\ref{pcl}) for classical probability we have to find the period (i.e. $2T$). Equivalently one can write $P_{Cl}(\xi)$ in the form:
\begin{equation}
P_{Cl}(\xi)=\frac{N_c}{|\eta(\xi)|},
\label{pclm}
\end{equation}
and find the unknown constant $N_c$ from the normalization:
\begin{equation}
\int\limits_{\xi_{min}}^{\xi_{max}} P_{Cl}(\xi)d\xi=1.
\label{norma}
\end{equation}
This is an elementary integral, from which one immediately gets 
\begin{equation}
N_c=\frac{\sqrt{-2{\cal E}}}{\pi}\, ,
\label{nn}
\end{equation}
independently on the value of $\beta$.

In the simplest case~(\ref{phi}), we now have
\begin{equation}
\phi(\xi,\xi')=\frac{\kappa\sqrt{-2{\cal E}}}{\pi}\,\chi_{\xi'}(\xi),
\label{phimorse}
\end{equation}
and the improved probability distribution $P(\xi)$ is given by~(\ref{prob}) with~(\ref{pphi}).

\begin{figure}[ht]
\centering
{\includegraphics[width=0.45\textwidth]{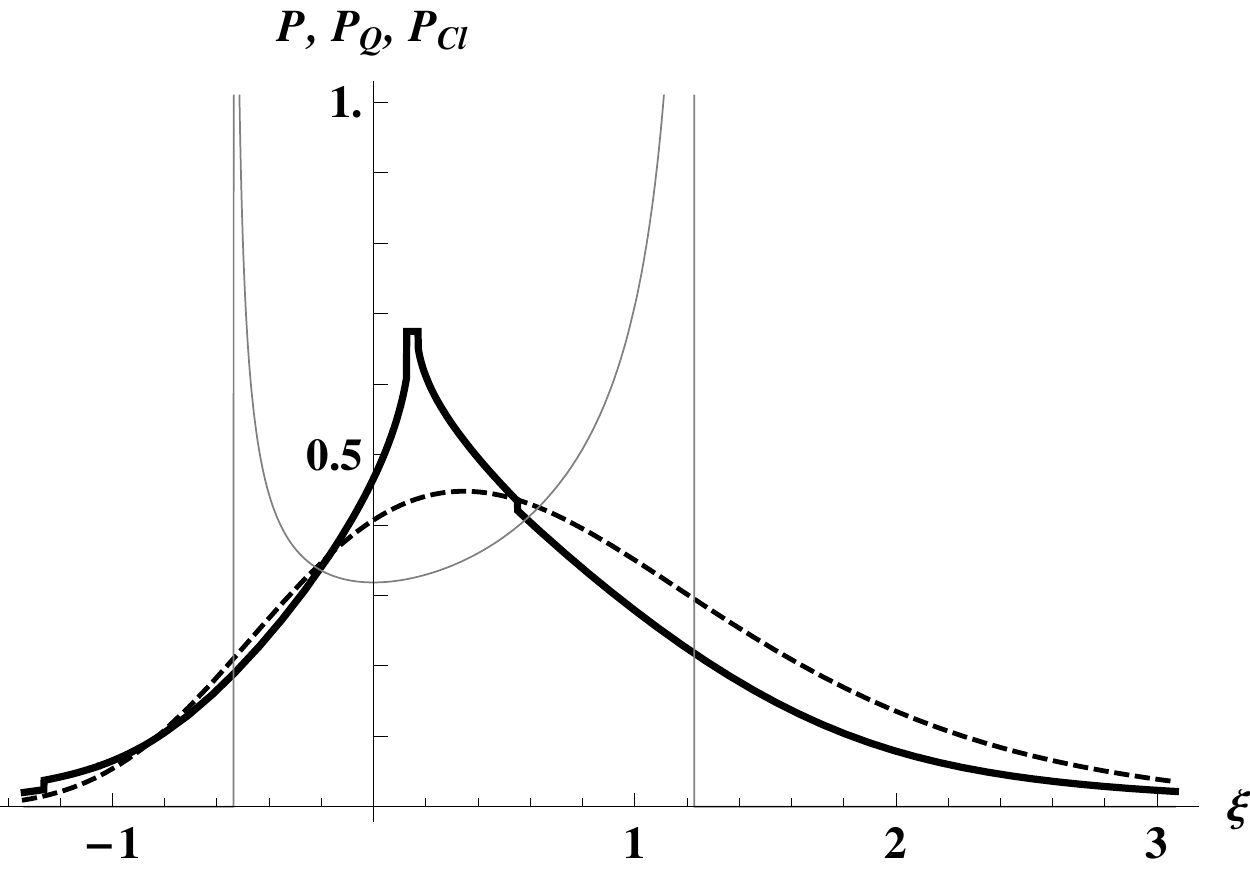}
\caption{Same as Fig~\ref{osc1}, but for the Morse potential. The parameter $\beta$ has been put to $1/2$.}
\label{morse1}}
\end{figure}

The quantum distribution for the Morse potential may be found elsewhere~\cite{land}:
\begin{equation}
P_Q(\xi)=N_q e^{-2\sqrt{2\beta}\, e^{-\xi}}e^{-2\sqrt{-2{\cal E}}\,\xi},
\label{pqmorse}
\end{equation}
where
\begin{equation}
{\cal E}=\frac{1}{2}\left(\frac{1}{2} - \sqrt{2\beta}\right)
\label{emorse}
\end{equation}
is the ground state dimensionless energy and $N_q$ can be obtained numerically (for any chosen value of $\beta$) from normalization condition (for instance for $\beta=1/2$ one gets $N_q\approx 3.00609$).

The results are shown in Fig.~\ref{morse1}. The value of the parameter $\beta$ has to be greater than $1/8$, since otherwise the potential has no bound states, and we have chosen $\beta=1/2$ which gives ${\cal E}=-1/4$.  We see that the classical distribution, after having included the uncertainty relation, agrees with the quantum distribution except for the small region around the maximal value of particle momentum. We are already familiar with this effect. As before, it may be cured by choosing one of the modified functions $\phi(\xi,\xi')$ shown in Fig.~\ref{heisen}.

\begin{figure}[t]
\centering
{\includegraphics[width=0.45\textwidth]{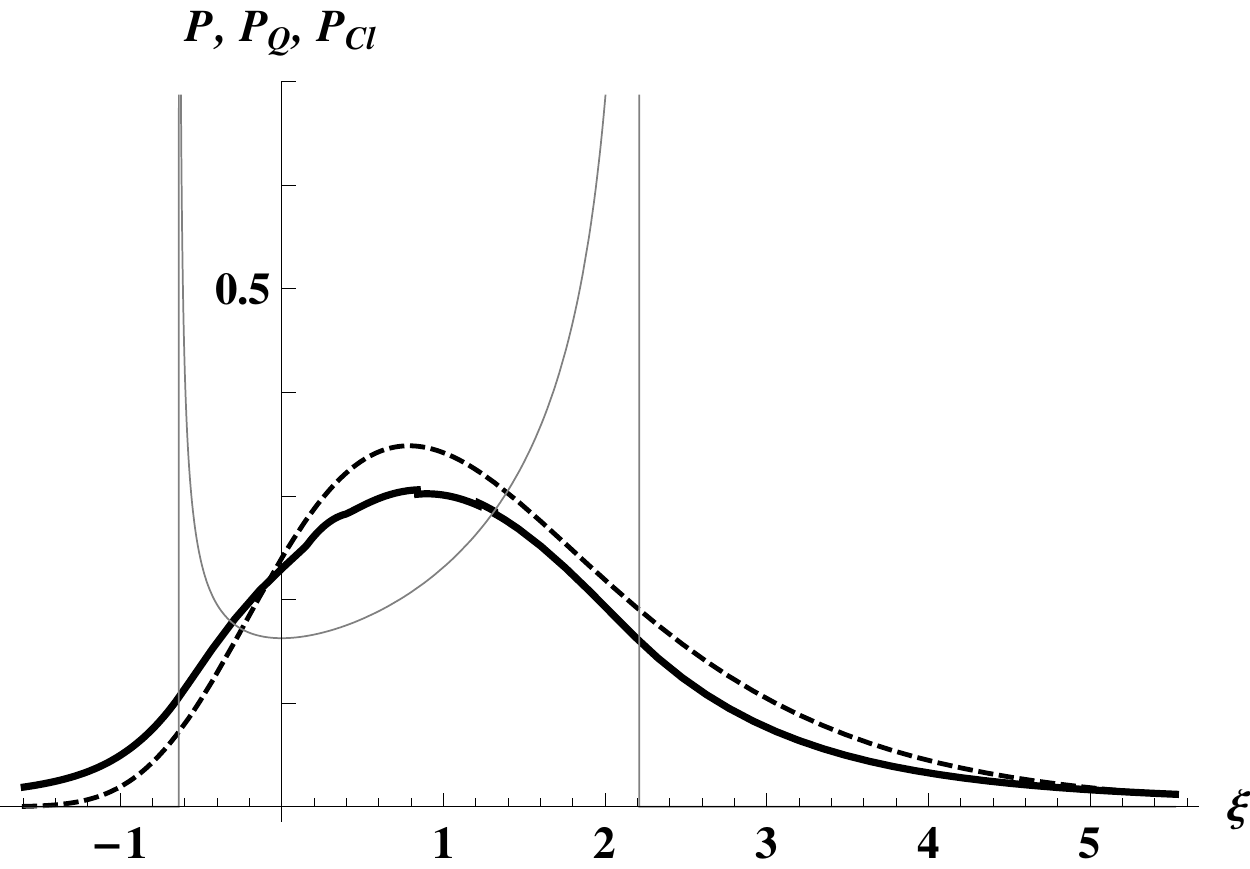}
\caption{Same as Fig.~\ref{morse1} but with the triangle form of $\phi(\xi,\xi')$.}
\label{morse2}}
\end{figure}

We first take the triangle form of $\phi(\xi,\xi')$. Since there are no unphysical rigid walls in the potential, no further modifications are required, and we can directly apply~(\ref{phi2}),~(\ref{Posc2}) and~(\ref{prob}) with the substitution~(\ref{morseeta}) for momentum. We expect $P(\xi)$ to approximate relatively well $P_Q(\xi)$ in the whole space. This is actually what happens as may be seen in Fig.~\ref{morse2}. The agreement becomes really satisfactory. 

\begin{figure}[h]
\centering
{\includegraphics[width=0.45\textwidth]{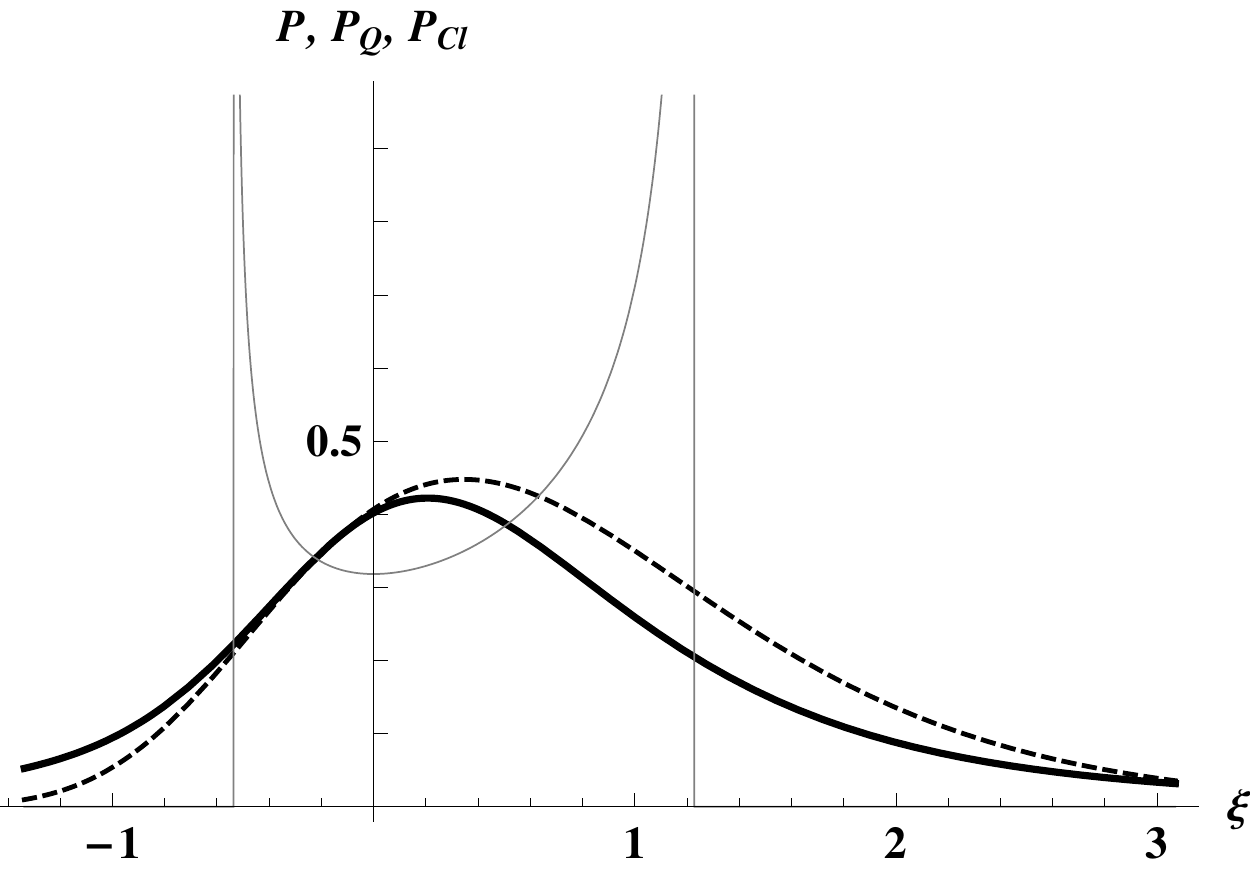}
\caption{Same as Fig.~\ref{morse1} but with the Gaussian form of $\phi(\xi,\xi')$.}
\label{morse3}}
\end{figure}

For Gaussian form of~ $\phi(\xi,\xi')$, we simply take~(\ref{phi3}) which leads to the plot in Fig.~\ref{morse3}. Just as we are already used to, the differences between the curves of the last two figures are minor.

\subsubsection{Some special potential}
\label{spec}
 
Our next testing example is some special potential, which --- up to our knowledge --- has no special name and which sometimes is viewed as resulting from the radial reduction of the spherically symmetric harmonic oscillator after having eliminated the angular variables~\cite{dere}.
The Hamiltonian of this model has the form
\begin{equation}
{\cal H}= \frac{p^2}{2m}+V_0\left(\frac{d}{x}-\frac{x}{d}\right)^2,
\label{specpot}
\end{equation}
with the condition $x>0$. The dimensionless variables remain the same as in the case of the Morse potential, and are given by~(\ref{mordim}). The same refers to the parameter $\beta$ defined in~(\ref{beta}).

In these variables the Eq.~(\ref{specpot}) becomes
\begin{equation}
\frac{\eta^2}{2}+\beta\left(\frac{1}{\xi}-\xi\right)^2={\cal E},
\label{specd}
\end{equation}
and for the momentum we get
\begin{equation}
|\eta(\xi)|=\sqrt{2({\cal E}-\beta(1/\xi -\xi)^2)}.
\label{momspec}
\end{equation}
The turning points are
\begin{subequations}\label{xsm}
\begin{align}
\xi_{min}&=\sqrt{1+{\cal E}/(4\beta)}-\sqrt{{\cal E}/(4\beta)},\,\label{xsm1}\\
\xi_{max}&=\sqrt{1+{\cal E}/(4\beta)}+\sqrt{{\cal E}/(4\beta)}.\label{xsm2}
\end{align}
\end{subequations}

In this example we do not have any rigid wall, nonetheless we cannot allow the uncertainty cells to 
extend to the negative values of $\xi$, which would be unphysical. Therefore we will use formulas~(\ref{Lgra}) and ~(\ref{phigrav}), together with the classical probability distribution in the form
\begin{equation}
P_{Cl}(\xi)=\frac{N_c}{|\eta(\xi)|}.
\label{pclaspec}
\end{equation} 
The constant $N_c$ may be explicitly obtained from the condition~(\ref{norma}), but this time it is too lengthy expression to be quoted below. Its numerical value for $\beta=2$ (which is chosen below as an exemplary value) is $N_c\approx 1.27324$. 

\begin{figure}[ht]
\centering
{\includegraphics[width=0.45\textwidth]{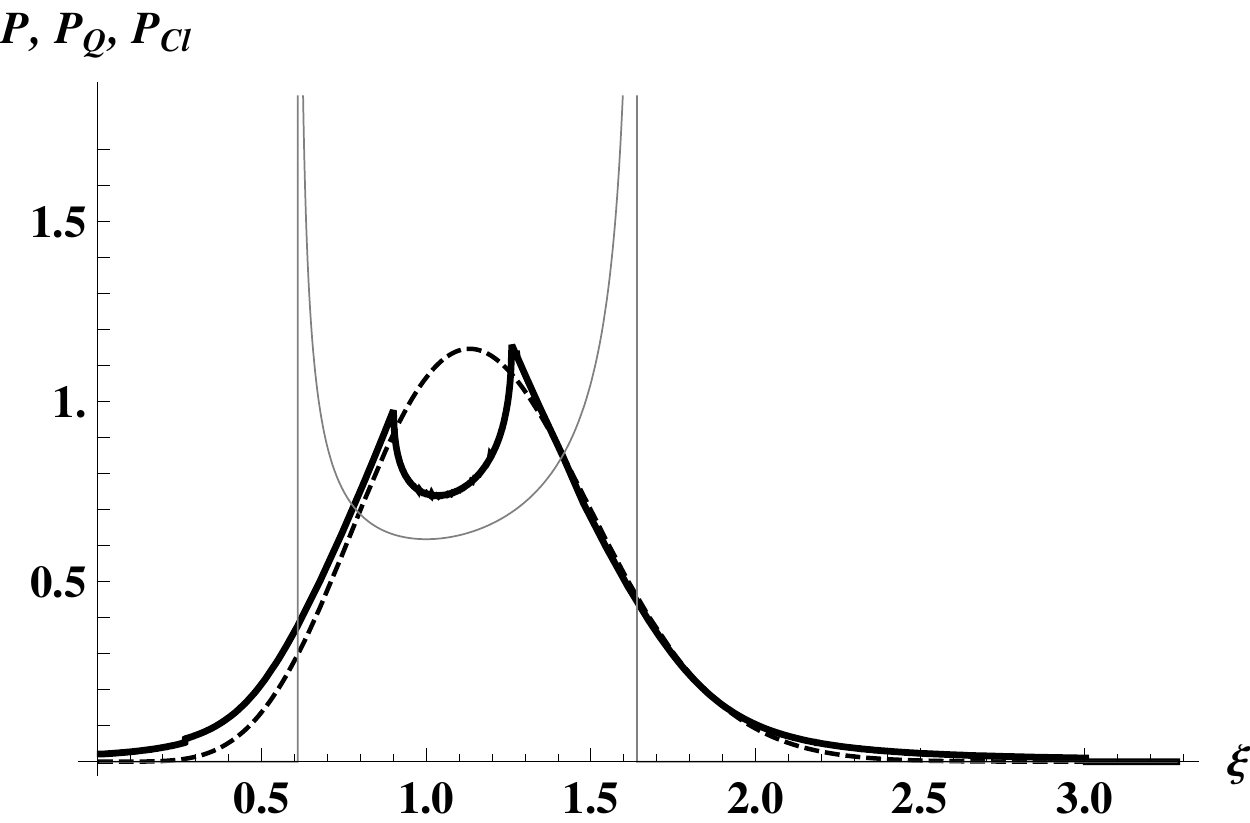}
\caption{Same as Fig.~\ref{osc1} but for the potential defined in~(\ref{specpot}). The parameter $\beta$ has been chosen to be equal to $2$.}
\label{bmw1}}
\end{figure}

The quantum ground state has the energy 
\begin{equation}
{\cal E} = \sqrt{2\beta}\left(1 + \sqrt{1/4+2\beta} - \sqrt{2\beta}\right),
\label{enbmw}
\end{equation}
and its probability distribution is~\cite{gold}
\begin{equation}
P_Q(\xi)=N_q\frac{e^{-\beta\xi^2}}{\xi^{1+\sqrt{1+8\beta}}},
\label{pqbmw}
\end{equation}
with the normalization constant given by
\begin{equation}
N_q=\frac{2\,\beta^{1 +\sqrt{1/4 + 2\beta}}}{\Gamma(1 + \sqrt{1/4 +2 \beta})}.
\label{nqbmw}
\end{equation}

The calculated probability distributions shown in Fig.~\ref{bmw1} confirm our earlier results: the agreement between $P(\xi)$ and $P_Q(\xi)$ is excellent except for a narrow strip around the minimum, which moreover can further be improved.

\begin{figure}[ht]
\centering
{\includegraphics[width=0.45\textwidth]{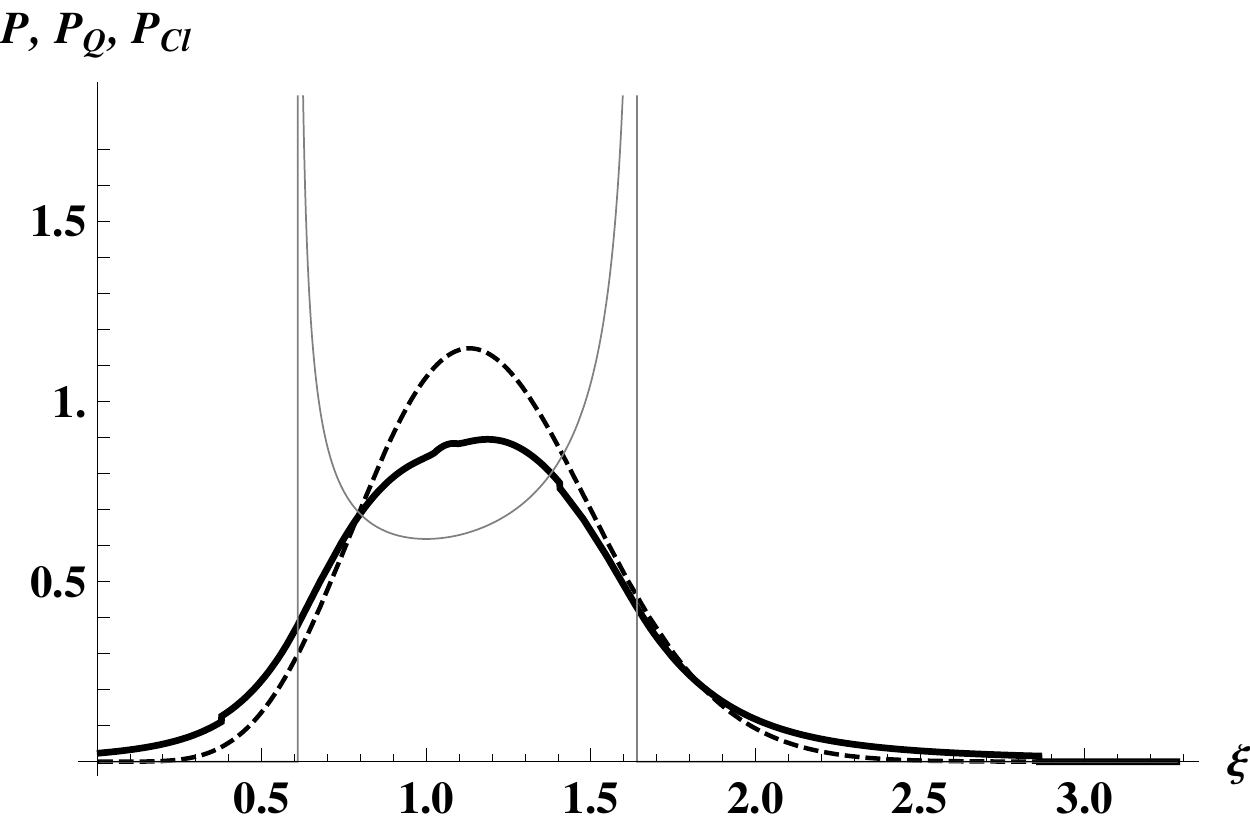}
\caption{Same as Fig.~\ref{bmw1}, but with the use of~(\ref{phigrav2}).}
\label{bmw2}}
\end{figure}

\begin{figure}[ht]
\centering
{\includegraphics[width=0.45\textwidth]{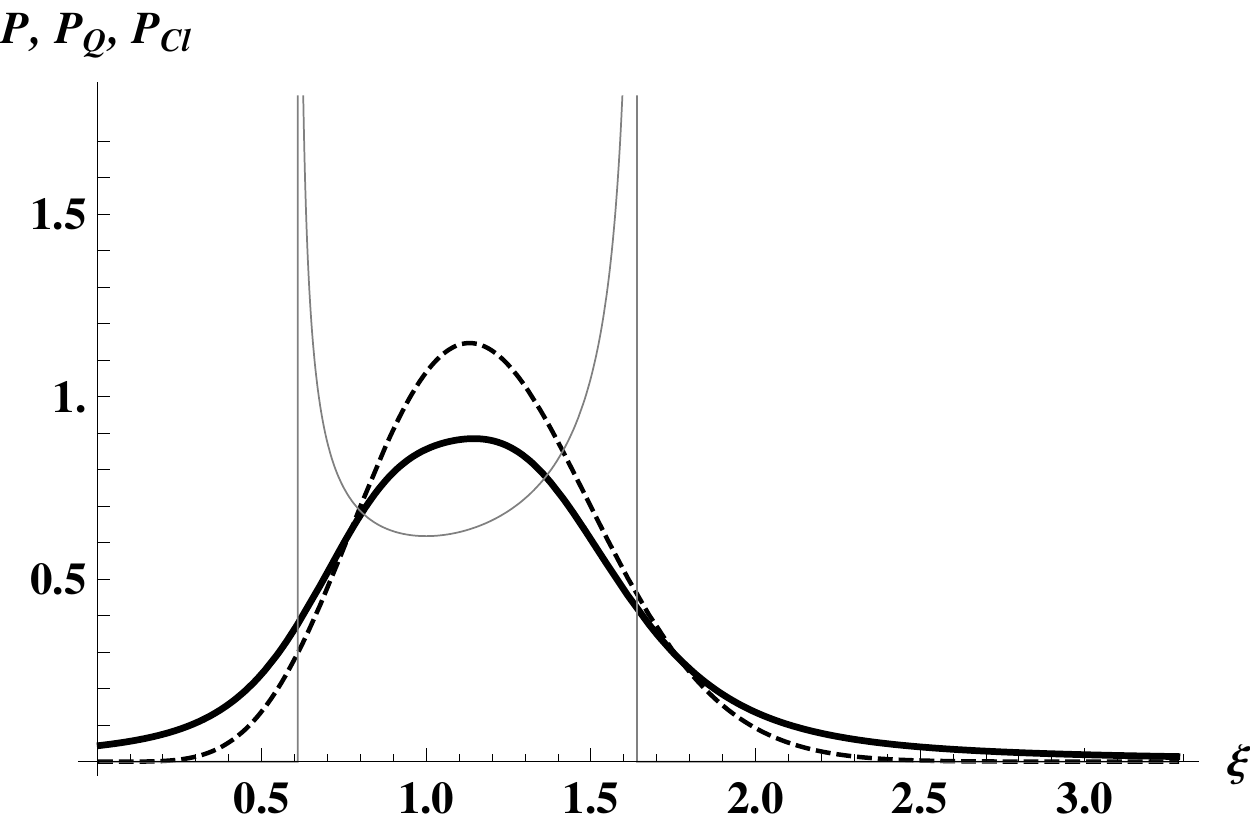}
\caption{Same as Fig.~\ref{bmw1}, but with the use of~(\ref{phigrav3}).}
\label{bmw3}}
\end{figure}

The results for more smooth forms of the function $\phi(\xi,\xi')$ may be immediately obtained with the use of formulas~(\ref{phigrav2}) and~(\ref{phigrav3}). They are plotted in Fig.~\ref{bmw2} and~\ref{bmw3}.

\subsection{More than one dimension}
\label{md}

In this subsection we apply the method to the motion in more than one dimension. As examples we will consider the hydrogen atom and the asymmetric two-dimensional harmonic oscillator.

\subsubsection{Hydrogen atom}
\label{hydro}

Due to the spherical symmetry of the problem (and of the ground state) it may be effectively reduced to the one-dimensional case and our procedure may be directly applied. The classical Hamiltonian in the case of zero angular momentum is
\begin{equation}
{\cal H}=\frac{p^2}{2m}-\frac{1}{4\pi\epsilon_0}\, \frac{e^2}{r}.
\label{Hh}
\end{equation}
If we introduce the Bohr radius
\begin{equation}
a_0=\frac{4\pi\epsilon_0\hbar^2}{me^2},
\label{bohr}
\end{equation}
we can define the dimensionless parameters in a way similar to~(\ref{mordim}):
\begin{subequations}\label{hdim}
\begin{align}
\xi&=\frac{x}{a_0}\, ,\;\;\;\;\; \eta=\frac{a_0}{\hbar}\, p\label{hdim1}\\
\tau&=\frac{\hbar}{ma_0^2}\, t,\;\;\;\;\; {\cal E}=\frac{a_0^2m}{\hbar^2}\,E,\label{hdim2}
\end{align}
\end{subequations}
which leads to the simple energy equation
\begin{equation}
\frac{\eta^2}{2}-\frac{1}{\xi}={\cal E},
\label{hdima}
\end{equation}
where ${\cal E}<0$. The turning points are
\begin{equation}
\xi_{min}=0,\;\;\;\;\; \xi_{max}=-\frac{1}{\cal E}\,.
\label{tuph}
\end{equation}

\begin{figure}[ht]
\centering
{\includegraphics[width=0.45\textwidth]{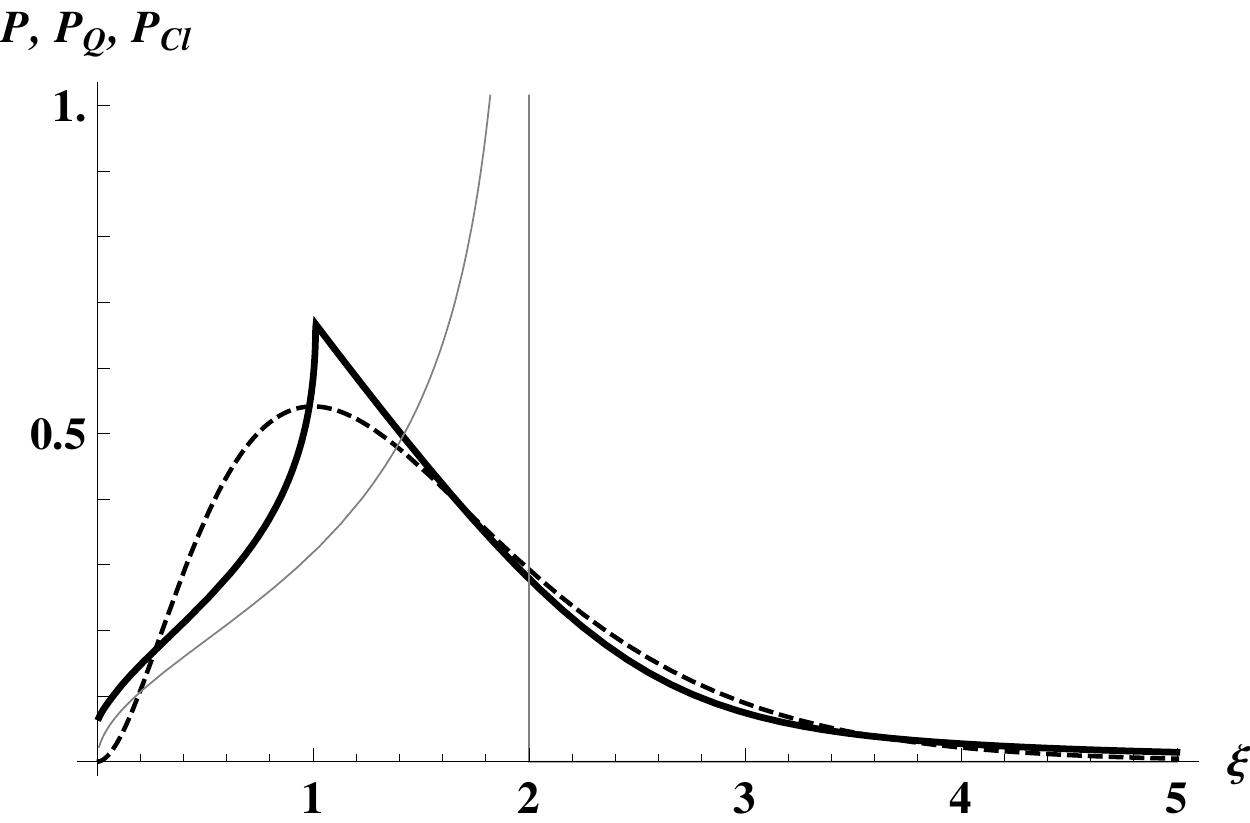}
\caption{Same as Fig.~\ref{grav1}, but for hydrogen atom.}
\label{hydrogen1}}
\end{figure}

The classical probability density is
\begin{equation}
P_{Cl}(\xi)=\frac{1}{T|\eta(\xi)|},
\label{pclh}
\end{equation}
where 
\begin{equation}
|\eta(\xi)|=\sqrt{2(1/\xi+{\cal E})}.
\label{etah}
\end{equation}
$T$ is, as usually, a half of the period. In the formula~(\ref{pclh}) it should be expressed in the dimensionless variables i.e. we have to put $T=\pi(-2{\cal E})^{-3/2}$.

To define the modified classical probability density we apply the same formulas as in subsection~\ref{gravi}. The region $\xi<0$ has to be excluded from the uncertainty cells, with the appropriate change in their normalization. We first take~(\ref{phigrav}) and obtain for $P(\xi)$ the behavior plotted in Fig.~\ref{hydrogen1}. For the energy we put ${\cal E}=-1/2$, which corresponds to the $1S$ state in the hydrogen atom. The quantum radial distribution for this state is well known to be~\cite{schiff}
\begin{equation}
P_Q(\xi)=4\xi^2 e^{-2\xi}.
\label{pqh}
\end{equation}

\begin{figure}[ht]
\centering
{\includegraphics[width=0.45\textwidth]{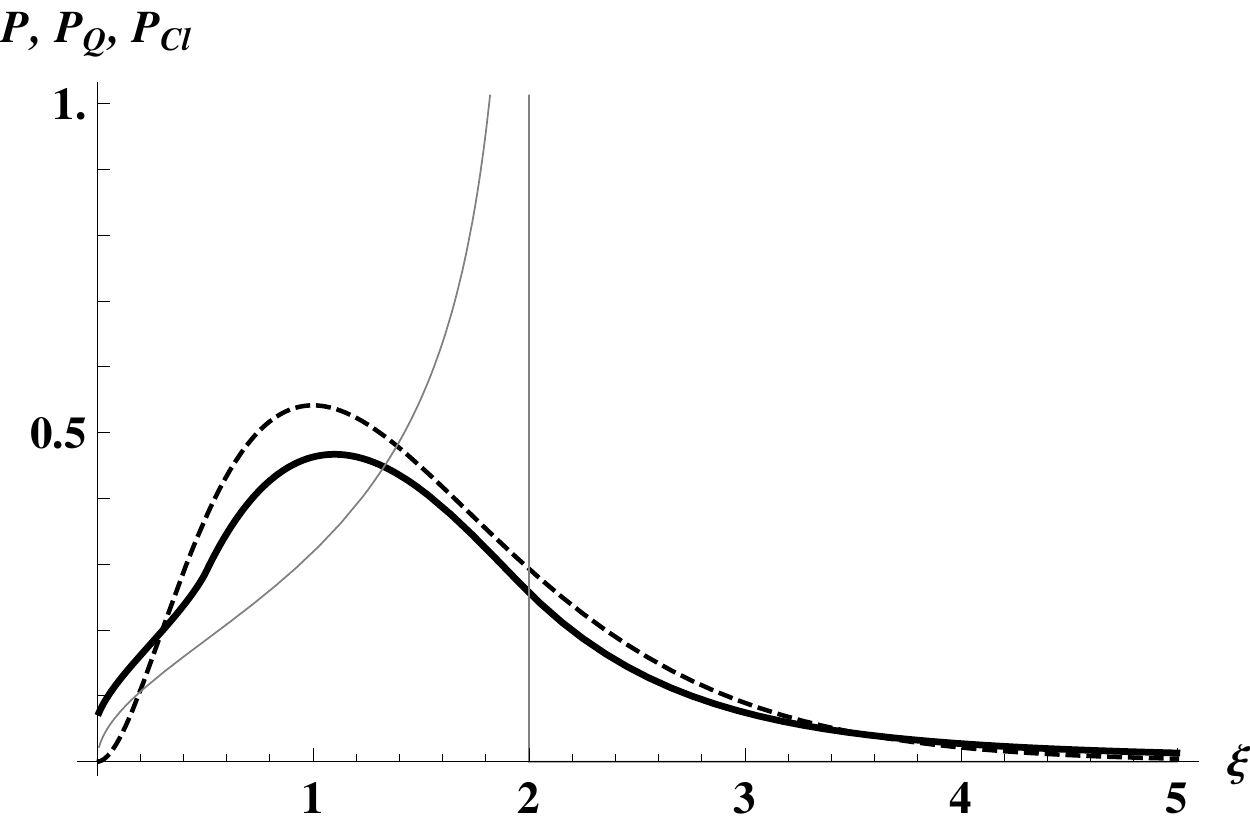}
\caption{Same as Fig.~\ref{grav2}, but for hydrogen atom.}
\label{hydrogen2}}
\end{figure}

\begin{figure}[ht]
\centering
{\includegraphics[width=0.45\textwidth]{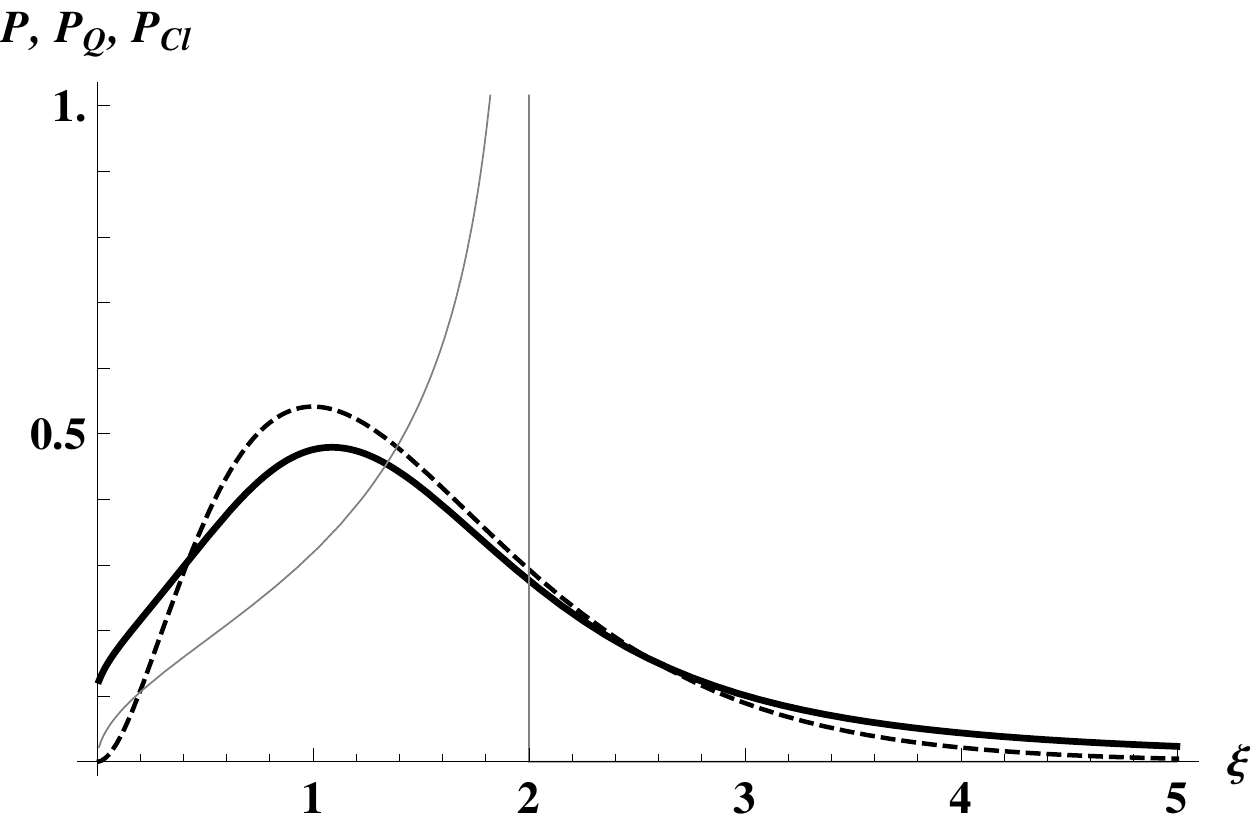}
\caption{Same as Fig.~\ref{grav3}, but for hydrogen atom.}
\label{hydrogen3}}
\end{figure}

The improved results obtained with the use of (\ref{phigrav2}) and~(\ref{phigrav3}), are given in Figs~\ref{hydrogen2} and~\ref{hydrogen3}. They confirm the conclusions from earlier plots: the general agreement with the quantum distributions and relative insensitivity to technicalities connected with the chosen classical probability distributions inside the uncertainty cell.

\subsubsection{Asymmetric harmonic oscillator}
\label{aharm}

Let us now consider the two-dimensional system with the Hamiltonian
\begin{equation}
{\cal H}=\frac{1}{2m}(p_x^2+p_y^2)+\frac{m}{2}(\omega_x^2x^2+\omega_y^2y^2),
\label{hoscas}
\end{equation}
with $\omega_x\neq\omega_y$. We choose that case to see, how the proposed method works in a system without rotational symmetry. It may also serve as a simple model of molecular vibrations.

The quantum probability density can be obtained by separation of variables in the Schr\"odinger equation and is a product of one-dimensional densities:
\begin{equation}
P_Q(x,y)=\frac{m\sqrt{\omega_x\omega_y}}{\hbar\pi}\,e^{-m\omega_xx^2/hbar}e^{-m\omega_yy^2/\hbar}.
\label{qas}
\end{equation}
In the dimensionless variables
\begin{subequations}\label{xixy}
\begin{align}
&\xi_x=\sqrt{\frac{m\omega_x}{\hbar}}\,x,\;\;\;\;\; \xi_y=\sqrt{\frac{m\omega_y}{\hbar}}\,y,\label{xixya}\\
&\eta_x=\frac{p_x}{\sqrt{m\hbar\omega_x}},\;\;\;\;\; \eta_y=\frac{p_y}{\sqrt{m\hbar\omega_y}},\label{xixyb}
\end{align}
\end{subequations}
it may be given a rotationally symmetric (in the plane $\xi_x\xi_y$) form
\begin{equation}
P_Q(\xi_x,\xi_y)=\frac{1}{\pi}\,e^{-\xi_x^2-\xi_y^2}.
\label{qasr}
\end{equation}

Let us define the auxiliary angle $\alpha$, which specifies the asymmetry of the potential
\begin{equation}
\sin\alpha= \frac{\omega_x}{\sqrt{\omega_x^2+\omega_y^2}},\;\;\;\;\;\; \cos\alpha= \frac{\omega_y}{\sqrt{\omega_x^2+\omega_y^2}}
\label{alpha}
\end{equation}
Considering classical motion of energy $E$, corresponding to that of the quantum ground state, we can define two dimensionless energies, which will be useful, and the corresponding dimensionless times:
\begin{subequations}\label{tet}
\begin{align}
&{\cal E}=\frac{E}{\hbar\sqrt{\omega_x^2+\omega_y^2}},\;\;\;\;\;\; \tilde{\cal E}=\frac{E}{\hbar\sqrt{\omega_x\omega_y}},
\label{teta}\\
&\tau=\sqrt{\omega_x^2+\omega_y^2}\, t,\;\;\;\;\;\; \tilde{\tau}=\sqrt{\omega_x\omega_y}\, t.\label{tetb}
\end{align}
\end{subequations}
Inserting the ground state energy $E=\hbar(\omega_x+\omega_y)/2$, one obtains ${\cal E}=(\sin\alpha+\cos\alpha)/2$ and $\tilde{\cal E}=(\tan\alpha+\cot\alpha)/2$. 
To find the classical probability density we have to know the particle velocity at each point of the classically allowed region. To this goal it is convenient to introduce new variables, as
\begin{subequations}\label{2da}
\begin{align}
&\tilde{\xi}_x=\xi_x\tan^{1/4}\alpha,\;\;\;\;\;\;\;\; \tilde{\xi}_y=\xi_y\cot^{1/4}\alpha\label{2da1}\\
&\tilde{\eta}_x=\eta_x\tan^{1/4}\alpha,\;\;\;\;\;\;\;\; \tilde{\eta}_y=\eta_y\cot^{1/4}\alpha.\label{2da2}
\end{align}
\end{subequations}
Eq.~(\ref{hoscas}) may now be given the form:
\begin{equation}
\frac{1}{2}(\tilde{\eta}_x^2+\tilde{\eta}_y^2+\tilde{\xi}_x^2+\tilde{\xi}_y^2)=\tilde{\cal E},
\label{hti}
\end{equation}
from which one gets
\begin{equation}
|\tilde{\eta}(\tilde{\xi},\tilde{\phi})|=\sqrt{2\tilde{\cal E}-\tilde{\xi}^2},
\label{soleta2}
\end{equation}
where $|\tilde{\eta}|=\sqrt{\tilde{\eta}_x^2+\tilde{\eta}_y^2}$, and $\tilde{\xi},\tilde{\phi}$ being the polar coordinates in the plane $\tilde{\xi}_x\tilde{\xi}_y$. It should be noted, that the transition from the initial Cartesian coordinates to the present ones leads to the Jacobian merely equal to a constant, which disappears when the probability is normalized to unity. 

Due to the symmetry of~(\ref{hti}) the classical probability density is $\tilde{\phi}$ independent, and therefore may be written as
\begin{equation}
P_{Cl}(\tilde{\xi},\tilde{\phi})=\frac{\tilde{C}}{|\tilde{\eta}(\tilde{\xi},\tilde{\phi})|},
\label{pcl2dt}
\end{equation}
where $\tilde{C}$ is certain constant to be fixed from the probability normalization condition. Coming back to the variables $\xi_x,\xi_y$, we obtain
\begin{align}
P_{Cl}&(\xi_x,\xi_y)=\nonumber\\
&\frac{C}{\sqrt{(2{\cal E} -\xi_x^2\sin\alpha-\xi_y^2\cos\alpha)(\xi_x^2\sin\alpha+\xi_y^2\cos\alpha)}},
\label{pcl2d}
\end{align}
where the additional factor in the denominator comes from the Jacobian. Integration over whole allowed region leads to
\begin{equation}
C=\frac{\sqrt{\sin\alpha\cos\alpha}}{\pi^2}.
\label{coc}
\end{equation}

\begin{figure}[ht]
\centering
{\includegraphics[width=0.45\textwidth]{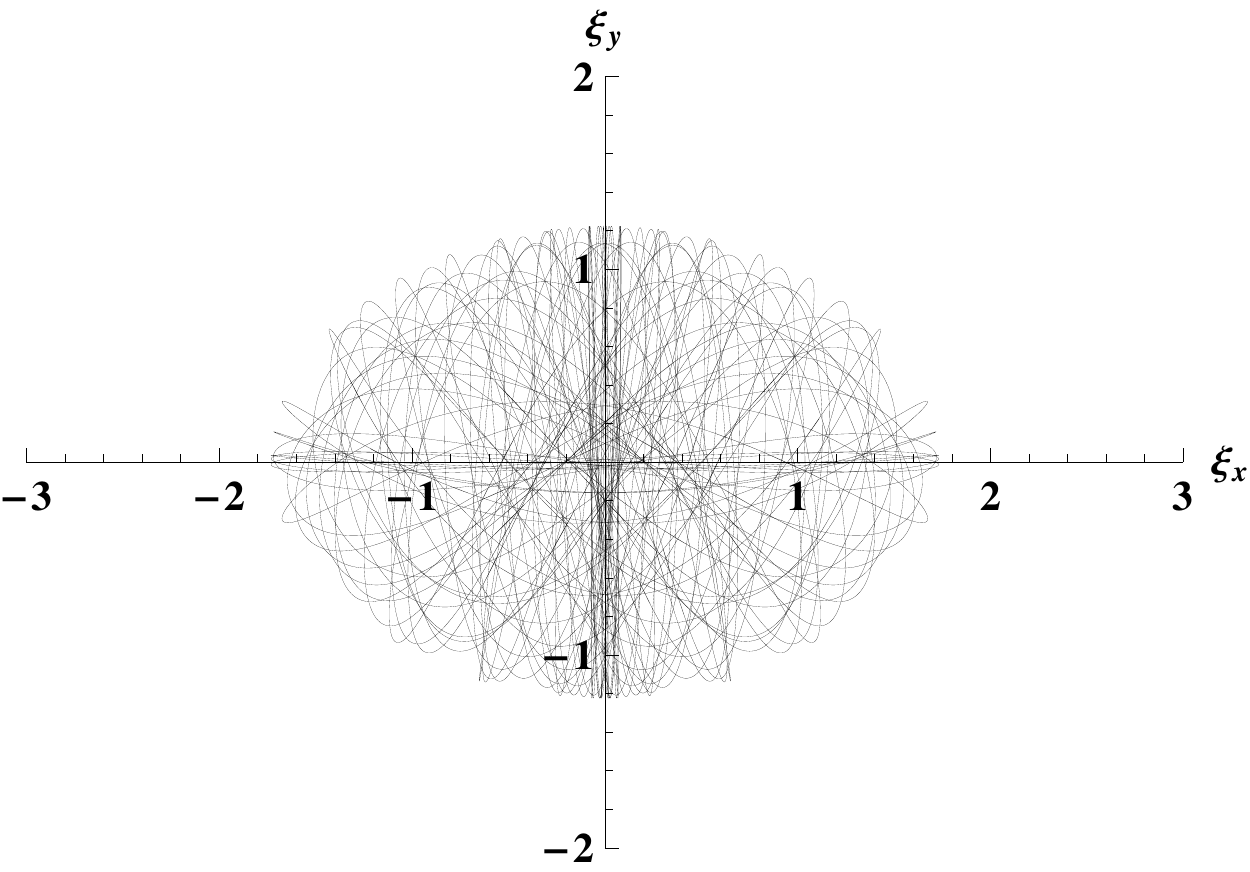}
\caption{The $50$ random classical trajectories for the fixed value of ${\cal E}=(\sin\alpha+\cos\alpha)/2$. The characteristic frequencies of the cavity have been chosen to be $\omega_y/\omega_x=2$.}
\label{tr2D}}
\end{figure}

\begin{figure*}[ht]
\centering
{\includegraphics[width=0.95\textwidth]{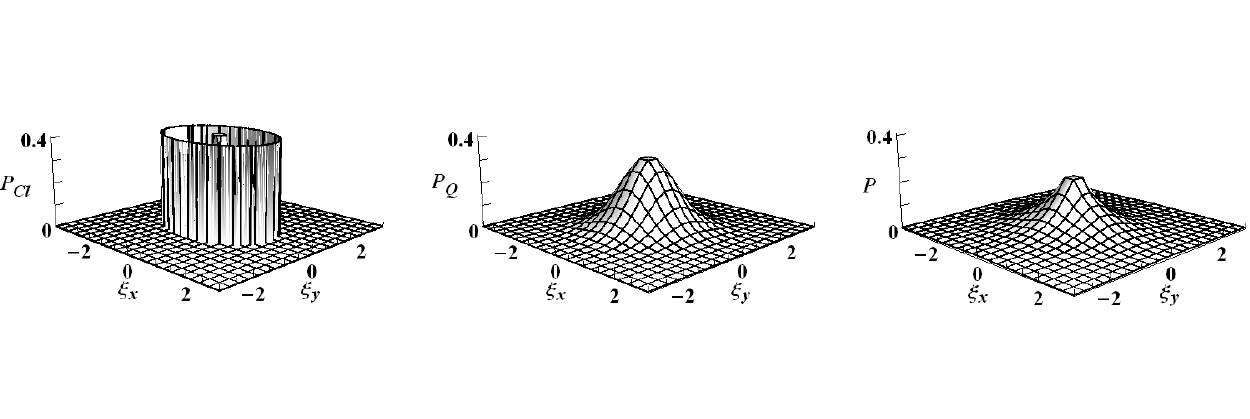}
\caption{The classical (first graph), quantum (second graph) and improved classical probability density distribution (third graph) for the ground state of the asymmetric harmonic oscillator obtained for $\kappa=1.8$ and $\omega_y/\omega_x=2$.}
\label{osc2D}}
\end{figure*}

It is obvious from~(\ref{pcl2dt}) that $P_{Cl}$ is rotationally invariant only in the plane $\tilde{\xi}_x\tilde{\xi}_y$, and not $\xi_x\xi_y$, where level sets are rather ellipses, and the major contribution for the probability comes from the vicinity of the curve $\tilde{\xi}^2=2\tilde{\cal E}$. The shape $P_{Cl}(\xi_x,\xi_y)$ is shown on the first plot of Fig.~\ref{osc2D}. The visible enhancement of the classical probability close to the origin is connected with the fact, that for various trajectories the particle approaches very often the cavity center and therefore it becomes more probable to be found there. This can be seen in Fig.~\ref{tr2D}, where the compilation of $50$ exemplary random trajectories is presented. In turn on this graph the increased probability on the perimeter is not exposed, since it is connected with the particle velocity and not its position.

The distribution of $P_{Cl}$ is in strong contradiction with the quantum probability $P_Q$ (shown on the second plot of Fig.~\ref{osc2D}), which has Gaussian character and in these variables is rotationally invariant.

Now we are in the position to apply our scheme to verify, whether it leads the (almost) quantum distribution. Due to the multidimensional nature of our problem, it requires some small modifications. First, to define the generalization of the function~(\ref{xipm}), we need two uncertainties $\Delta\xi_x$ and $\Delta\xi_y$. They are connected with the velocities $\eta_x$ and $\eta_y$ at a given point, but these are unknown. From the energy equation~(\ref{hti}) we obtain only $\sqrt{\tilde{\eta}_x^2+\tilde{\eta}_y^2}$, but the direction of the motion --- and simultaneously $\Delta\xi_x$ and $\Delta\xi_y$ at each point --- is unspecified and depends on the initial conditions. Therefore we introduce a certain angle $\beta$, which defines this direction. For any fixed $\beta$ the values of $\Delta\xi_x$ and $\Delta\xi_y$ become known and the final probability distribution will be that obtained by averaging over possible directions. This angle averaging refers to the initial state of the classical particle only and should not be confused with the directions in the harmonic potential: the trap does not change it shape as given by~(\ref{hoscas}).
In this way we have
\begin{subequations}\label{pre}
\begin{align}
|\eta_x(\xi_x,\xi_y)|&=\sqrt{2{\cal E} -\xi_x^2\sin\alpha-\xi_y^2\cos\alpha}\, \cot^{1/4}\alpha\cos\beta,\label{prea}\\
|\eta_y(\xi_x,\xi_y)|&=\sqrt{2{\cal E} -\xi_x^2\sin\alpha-\xi_y^2\cos\alpha}\, \tan^{1/4}\alpha\sin\beta,\label{preb}
\end{align}
\end{subequations}
and we can define the generalization of the function $\phi$ given by~(\ref{phi}):
\begin{align}
\phi(\bm{\xi},\bm{\xi}')&=(\Delta\xi_x\Delta\xi_y)^{-1}\chi_{x,\bm{\xi}'}(\xi_x)\chi_{y,\bm{\xi}'}(\xi_y)\nonumber\\
&=\kappa^2|\eta_x(\bm{\xi}')\eta_y(\bm{\xi}')|\chi_{x,\bm{\xi}'}(\xi_x)\chi_{y,\bm{\xi}'}(\xi_y).
\label{phi2d}
\end{align}
As before, $\chi_{x,\bm{\xi}'}(\xi_x)$ is the characteristic function of the interval
\begin{equation}
D_{x,\bm{\xi}'}=[\xi_{x-}, \xi_{x+}],
\label{interval1}
\end{equation}
where
\begin{equation}
\xi_{x\pm}=\xi_x'\pm\frac{1}{2\kappa|\eta_x(\bm{\xi}')|}=\xi_x'\pm \frac{\Delta\xi_x}{2},
\label{xipm1}
\end{equation}
and similarly for $\chi_{y,\bm{\xi}'}(\xi_y)$. Now the improved probability density amplitude may be defined by the simple generalization of the formulas~(\ref{pphi}) and~(\ref{prob}):
\begin{equation}
\Phi(\bm{\xi},\bm{\xi}')=\phi(\bm{\xi},\bm{\xi}') P_{Cl}(\bm{\xi}').
\label{pphi1}
\end{equation}
and
\begin{equation}
P(\bm{\xi})=\langle\int\limits_{D_{Cl}}\Phi(\bm{\xi},\bm{\xi}')d\xi_x'd\xi_y'\rangle_\beta,
\label{prob1}
\end{equation}
where by $D_{Cl}$ we denoted the classically allowed region, i.e. the ellipse defined by the condition $\xi_x'^2\sin\alpha+\xi_y'^2\cos\alpha<2{\cal E}$.

If we preferred a Gaussian distribution of probability within the uncertainty cell, we would use, instead of~(\ref{pphi1}), the obvious generalization of~(\ref{Posc3}):
\begin{align}
\Phi(\bm{\xi},\bm{\xi}')&=\frac{2\kappa^2}{\pi}\, |\eta_x(\bm{\xi}')\eta_y(\bm{\xi}')|\label{Posc31}
\\ &\times e^{-2\kappa^2\eta_x(\bm{\xi}')^2(\xi_x-\xi_x')^2}e^{-2\kappa^2\eta_y(\bm{\xi}')^2(\xi_y-\xi_y')^2}P_{Cl}(\bm{\xi}').\nonumber
\end{align}

In Fig.~\ref{osc2D} we draw the improved probability distribution $P(\xi_x,\xi_y)$ obtained with the use of~(\ref{Posc31}), for the value of parameter $\kappa=1.8$ (this gives slightly better agreement than the `one-dimensional value' $1.7$) and compare it to the quantum probability $P_Q(\xi_x,\xi_y))$. The obtained result turns out again to be noteworthy. The main features and even the rotationally invariant character of $P_Q$  in the variables $\xi_x,\xi_y$ have been approximately restored from~(\ref{pcl2d}) thanks to the application of the uncertainty relation.

\section{Summary}
\label{sum} 

In the present paper we proposed a simple way to improve the classical probability density distribution, by incorporating the Heisenberg uncertainty principle in order to make it consistent with the quantum distribution.  It is based on the picture of a wave packet bouncing between classical turning points. We concentrated on the ground states for which the discrepancy between classical and quantum results was the most severe. The procedure was tested for various one-dimensional quantum-mechanical models: harmonic oscillator, linear (gravitational) potential, Morse potential and some special potential, as well as for models in more than one dimension: the hydrogen atom and asymmetric harmonic oscillator. They all show the unexpectedly good agreement between the improved distribution and the quantum one. This is all the more noteworthy that the proposed modification is very simple if not trivial.

In the region of large particle momenta, the agreement is not so good, but it was shown, that further improvements may be easily achieved by the slight and still simple modification of the classical probability distribution within the uncertainty cell. After this modification the obtained results are satisfactory for all tested potentials and for the whole space.

The comparison of all plots reveals one feature in common: the probability is slightly underestimated in the center of cavity and slightly overestimated far from it. This is a consequence of our simple assumption~(\ref{uncpos}) and entails a little larger values for the variance
\begin{equation}
\sigma_\xi^2=\langle (\xi-\langle\xi\rangle)^2\rangle,
\label{six}
\end{equation}
than those found from quantum distribution. In some small limits~(\ref{six}) may be modified by the choice of parameter $\kappa$, whose value (mostly $\kappa=1.7$) was fixed to obtain the {\em visual} agreement between $P(\xi)$ and $P_Q(\xi)$.

One might have a hope that for the determination of the approximate probability distribution one does not necessarily have to go through the often complicated process of finding the quantum state, but sufficient is the knowledge of the classical motion and of the uncertainty principle. The solving of ordinary differential equations describing the classical motion of a particle should be numerically much more efficient than that of the partial Schr\"odinger differential equation, for instance with the use of the Monte Carlo methods.

The presented idea seems promising and should be tested for more complicated cases, also going beyond the simple non-relativistic quantum mechanics. What is required, is the knowledge of velocities or momenta at each classically allowed point, which, at least at the numerical level, should be available. At the same time the method confirms the fundamental role played in quantum mechanics by the uncertainty principle. The comparison of classical and quantum results allows for better understanding of the underlying physics and especially the connection between quantum mechanics and the macroscopic world.

\section*{Acknowledgements}
I would like to thank to Prof. Usha Devi for pointing me to the interesting literature on this subject.


\begin{thebibliography}{99}
\bibitem{burk} C.E. Burkhardt and J.J Leventhal, {\em Foundations of Quantum Physics}, Springer, New York 2008.
\bibitem{bohr} N. Bohr, Z. Phys. {\bf 2}, 423(1920).
\bibitem{hol} P.R. Holland, {\em The quantum theory of motion}, Cambridge University Press, Cambridge 1993.
\bibitem{jammer} M. Jammer, {\em The Conceptual Development of Quantum Mechanics}, McGraw Hill Book Co., New York 1966.
\bibitem{darg} O. Darrigol,  Riv. Internaz. di Storia della Scienza {\bf 34}, 545(1997). 
\bibitem{robin} R.W. Robinett, Am. J. Phys. {\bf 63}, 823(1995).
\bibitem{devi} A.R. Usha Devi and H.S. Karthik, Am. J. Phys. {\bf 80}, 708(2012).
\bibitem{bdr} M. Belloni, M.A. Doncheski and R.W. Robinett, Physica Scripta {\bf 71}, 136(2005).
\bibitem{brandt} S. Brandt and H.D. Dahmen {\em The Picture Book of Quantum Mechanics}, Wiley, New York 1985.
\bibitem{sax} D.S. Saxon, {\em Elementary Quantum Mechanics}, McGraw-Hill, New York, 1968.
\bibitem{mil} W.H. Miller, J. Chem. Phys. {\bf 54}, 5386(1971).
\bibitem{schr} E. Schr\"odinger, Die Naturwissenschaften {\bf 14}, 644(1926).
\bibitem {lj} J. Lekner and H. Nguyen, Eur. J. Phys. {\bf 30}, L67(2009).
\bibitem{eh} P. Ehrenfest, Zeitschrift f\"ur Physik {\bf 45}, 455 (1927).
\bibitem{schiff} For instance L. Schiff, {\em Quantum Mechanics}, MC-Graw-Hill, New York 1968.
\bibitem{wen} G. Wentzel. Z. Physik, {\bf 38}, 518(1926).
\bibitem{kra} H.A. Kramers, Z. Physik, {\bf 39}, 828(1926).
\bibitem{bri} L. Brillouin, Compt. Rend. {\bf 183}, 24(1926).
\bibitem{gibbs} R.L. Gibbs, Am. J. Phys. {\bf 43}, 25 (1975)
\bibitem{nes} V. Nesvizhevsky {\em et al.},  Nature {\bf 415}, 297(2002).
\bibitem{liu} Q.-h. Liu, Front. Phys. {\bf 2}, 273(2007). 
\bibitem{pedram} P. Pedram, Eur. Phys. J. {\bf C 73}, 2609(2013).
\bibitem{abra} M. Abramowitz and I.A. Stegun, {\em Handbook of Mathematical Functions with Formulas, Graphs, and Mathematical Tables}, New York 1972, 9th ed.
\bibitem{flugge} S. Fl\"ugge, {\em Rechmetoden der Quantentheorie}, Springer, Berlin 1999.
\bibitem{berb} M.N. Berberan-Santos, E.N. Bodunov and L. Pogliani, J. Mat. Chem. {\bf 37}, 101(2004).
\bibitem{yoder} G. Yoder, Am. J. Phys. {\bf 74}, 404 (2006).
\bibitem{morse} P.M. Morse, Phys. Rev. {\bf 34}, 57(1929).
\bibitem{atkins} P. Atkins and R. Friedman, {\em Molecular Quantum Mechanics}, Oxford University Press, New York 2005. 
\bibitem{land} For instance L.D. Landau and E.M. Lifshitz, {Quantum Mechanics. Non-Relativistic Theory}, Pergamon, Oxford 1977, 3rd ed.
\bibitem{dere} J. Derezi\'nski and M. Wrochna, Annales Henri Poincar\'e {\bf 12}, 397(2011).
\bibitem{gold} I.I. Gold'man, V.D. Krivchenkov, V.I. Kogan, V.M. Galitskii, {\em Problems in Quantum Mechanics}, Infosearch Ltd., London 1960.
\bibitem{robik} R. Robinett, {\em Quantum Mechanics,Classical Results, Modern Systems, and Visualized Examples}, Oxford University Press, New York 2006.
\end{thebibliography}
\end{document}